\documentclass[letterpaper,12pt]{article}
\usepackage{tabularx} % extra features for tabular environment
\usepackage{amsmath,bm}  % improve math presentation
\usepackage{float}
\usepackage{indentfirst}
\usepackage{graphicx} % takes care of graphic including machinery
\usepackage[margin=1in,letterpaper]{geometry} % decreases margins
\usepackage{caption}
\usepackage{cite} % takes care of citations
\usepackage[colorlinks,linkcolor=magenta,anchorcolor=cyan,citecolor=blue]{hyperref} % adds hyper links inside the generated pdf file
\usepackage{tensor}
\hypersetup{
	colorlinks=true,       % false: boxed links; true: colored links
	linkcolor=blue,        % color of internal links
	citecolor=blue,        % color of links to bibliography
	filecolor=magenta,     % color of file links
	urlcolor=blue
}

\def\be{\begin{equation}}
\def\ee{\end{equation}}
\def\ba{\begin{eqnarray}}
\def\ea{\end{eqnarray}}

              \def\.{\cdot}

\begin{document}
\begin{center}

	\vspace{10pt}
	\large{\bf{Generalized proof of the linearized second law in general quadric corrected Einstein-Maxwell gravity}}
	
	\vspace{15pt}
	Xin-Yang Wang $^\text{a}$, Jie Jiang $^\text{b}$
	
	\vspace{15pt}
	\small{\it College of Education for the Future, Beijing Normal University, Zhuhai 519087, China}
	
	\vspace{30pt}
\end{center}
\begin{abstract}
	Although the entropy of black holes in any diffeomorphism invariant theory of gravity can be expressed as the Wald entropy, the issue of whether the entropy always obeys the second law of black hole thermodynamics remains open. Since the non-minimal coupling interaction between gravity and the electromagnetic field in the general quadric corrected Einstein-Maxwell gravity can sufficiently influence the expression of the Wald entropy, we would like to check whether the Wald entropy of black holes in the quadric corrected gravity still satisfies the second law. A quasistationary accreting process of black holes is first considered, which describes that black holes is perturbed by matter fields and eventually settle down to a stationary state. Two assumptions that the matter fields should obey the null energy condition and that a regular bifurcation surface exists on the background spacetime are further proposed. According to the two assumptions and the Raychaudhuri equation, we demonstrate that the Wald entropy monotonically increases along the future event horizon under the linear order approximation of the perturbation. This result indicates that the Wald entropy of black holes in the quadric corrected gravity strictly obeys the linearized second law of thermodynamics.
\end{abstract}
	\vfill {\footnotesize ~\\ $^\text{a}$ xinyangwang@bnu.edu.cn \\ $^\text{b}$ Corresponding author. jiejiang@mail.bnu.edu.cn}
\newpage

\section{Introduction}
Black holes are special spacetime structures predicted by General Relativity. The boundary of black holes is called the event horizon which is a null hypersurface. Pioneering work by Hawking \cite{Hawking:1971tu} demonstrated that the area of the event horizon never decreases along the direction of the time evolution. From this perspective, Bekenstein \cite{Bekenstein:1973ur} first suggested that there may be a strong connection between the area of the event horizon and the entropy of black holes. Using the quantum field theory in curved spacetime, Hawking \cite{Hawking:1974sw} showed that the temperature and the entropy of black holes are defined by its surface gravity and area of the event horizon. While the entropy of black holes is expressed as $S_{BH} = A / 4$, where $A$ is the area of the event horizon, and it is called the Bekenstein-Hawking entropy. From the definition of the temperature and the entropy of black holes, the four laws of thermodynamics of black holes are constructed, and black holes can be identical with adiabatic systems in the thermodynamics \cite{Bekenstein:1972tm, Bardeen:1973gs, Unruh:1976db}. In the four laws of thermodynamics, the two profound laws for black holes are the first and the second laws respectively. If we recognize black holes as thermodynamic systems, the two laws should be satisfied first. A natural question of whether the laws of thermodynamics are the robust feature for black holes in any covariant gravitational theory is proposed. Starting with this question, Iyer and Wald \cite{Wald:1993nt, Iyer:1994ys} established the first law of thermodynamics for any diffeomorphism invariant gravitational theory, and the entropy of black holes in the gravitational theory can be expressed as
\begin{equation}
	S_{\text{W}} = - 2 \pi \int_s d^n y \sqrt{\gamma} \frac{\partial \mathcal{L}}{\partial R_{abcd}} \boldsymbol{\epsilon}_{ab} \boldsymbol{\epsilon}_{cd}\,,
\end{equation}
where $\mathcal{L}$ is the Lagrangian of the theory of gravity, $\boldsymbol{\epsilon}_{ab}$ is the binormal for any slice of the event horizon, $\sqrt{\gamma}$ is the volume element of the cross section of the horizon, and $y$ is used to label the transverse coordinates on any slice. This entropy is called the Wald entropy and it is no longer proportional to the area of the event horizon. When a diffeomorphism invariant gravitational theory degraded into the Einstein theory, the form of Wald entropy is identical with Bekenstein-Hawking entropy as well.

In the study of the quantization of gravity, it still has not an appropriate scheme to quantize the gravitational field until now. To study the interaction between gravity and matter fields on the quantum scale, we should find the corresponding low-energy effective field theory (EFT) first. The low-energy EFT makes that some quantum corrected terms are added in the expression of the Lagrangian. These terms contain the higher curvature terms and the non-minimal coupling terms between gravity and matter fields. The existence of the quantum correction terms will change the expression of the Wald entropy substantially. Although it has been proved that the Wald entropy always obeys the first law of thermodynamics for any diffeomorphism invariant theory \cite{Wald:1993nt, Iyer:1994ys}, whether the Wald entropy in the theory of gravity with quantum correction still satisfies the second law of thermodynamics has not been generally demonstrated. Along this line of thought, the second law of black holes in the gravitational theory with higher curvature correction has been investigated in previous research works. According to the field redefinition, the Wald entropy of black holes in $f(R)$ gravity has been proved it satisfies the second law of thermodynamics \cite{Jacobson:1993vj, Jacobson:1995uq}. However, in the gravitational theory, the result shows that the second law is violated when two black holes emerge \cite{Sarkar:2010xp}. Bhattacharjee et al. \cite{Bhattacharjee:2015yaa} suggest that it is enough for us to examine the second law of black holes in quantum corrected gravitational theory under the linear order approximation of the process considering a quasistationary accretion process of black holes. Based on this perspective, the linearized second law of black holes in Gauss-Bonnet and Lovelock theory has been examined \cite{Chatterjee:2011wj, Kolekar:2012tq}. Furthermore, Wall \cite{Wall:2015raa} proposes a general method to check the linearized second law in arbitrary higher curvature gravity, while the expression of the entropy obeys the linearized second law is also given.

Although the theory of gravity with any higher curvature correction has been studied adequately, the situation that the Lagrangian contains the non-minimal coupling interaction between gravity and matter fields is not involved. Since the non-minimal coupling terms between the Riemann tensor and the scalar field are contained in the Lagrangian of the Horndeski gravity, we have examined the Wald entropy of black holes in the theory of gravity always satisfies the linearized second law under the first-order approximation of the quasistationary accretion process \cite{Wang:2020svl}. However, only studying the interaction between gravity and the scalar field cannot sufficiently examine whether the Wald entropy generally obeys the linearized second law in the theory of gravity with any non-minimal coupling matter fields. We should continue to consider the gravitational theory with other types of non-minimal coupling matter fields to examine the second law of black holes. Therefore, in the following, we would like to consider the most general quadric corrected Einstein-Maxwell theory of gravity. In this theory, expect the pure Einstein-Maxwell action, the quadric corrected terms also involved in the expression of the Lagrangian. These quadric correction terms mainly contain the self-interaction of the electromagnetic field, the non-minimal coupling interaction between gravity and the electromagnetic field, and the quadratic curvature correction. Similar to the case of higher curvature gravity, the non-minimal coupling terms also affect the expression of Wald entropy in the same way. Therefore, investigating whether the Wald entropy of black holes in the general quadric corrected gravitational theory still satisfies the linearized second law is also an important segment to examine the validity of the second law of black hole thermodynamics at the first-order approximation of the matter field perturbation.

The organization of the paper is as follows. In Sec. \ref{sec2}, the general quadric corrected Einstein-Maxwell gravity is introduced, and the Wald entropy of black holes in the theory of gravity is given. In Sec. \ref{sec3}, considering an quasistationary accreting process of black holes and assuming that the matter fields accreted by black holes should satisfy the null energy condition, we examine whether the Wald entropy still satisfies the linearized second law of black holes thermodynamics under the linear order approximation of the perturbation. The paper ends with discussions and conclusions in Sec. \ref{sec4}.

\section{General quadric corrected Einstein-Maxwell gravity and Wald entropy}\label{sec2}

We will consider the ($\text{n}+2$)-dimensional Einstein-Maxwell gravity with general quadric correction terms. These correction terms mainly contain the self-interaction terms of the electromagnetic field, the non-minimal coupling interaction between gravity and the electromagnetic field, and the quadratic curvature correction. The Lagrangian ($\text{n}+2$)-form of the gravitational theory can be formally written as
\begin{equation}\label{totallagrangian}
	\boldsymbol{\mathcal{L}} =\boldsymbol{\mathcal{L}}_{\text{EM}} + \Delta \boldsymbol{\mathcal{L}} + \boldsymbol{\mathcal{L}}_{\text{mt}}\,.
\end{equation}
The first part of Eq. (\ref{totallagrangian}) represents the Lagrangian of the pure Einstein-Maxwell gravitational theory,
\begin{equation}
	\boldsymbol{\mathcal{L}}_{\text{EM}} = \frac{1}{16 \pi} \left(R - F_{ab}F^{ab} \right) \boldsymbol{\epsilon}\,,
\end{equation}
where $R$ is the Ricci scalar, $\boldsymbol{F} = d \boldsymbol{A}$ is the electromagnetic strength, $\boldsymbol{A}$ is the vector potential of the electromagnetic field, and $\boldsymbol{\epsilon}$ is the volume element of the spacetime. The second part of Eq. (\ref{totallagrangian}) is the portion of the quadric correction, which can be decomposed into two parts, i.e.,
\begin{equation}\label{correctterm}
	\Delta \boldsymbol{\mathcal{L}} = \Delta \boldsymbol{\mathcal{L}}_1 + \Delta \boldsymbol{\mathcal{L}}_{\text{QC}}\,.
\end{equation}
The first part of Eq. (\ref{correctterm}) involves the non-minimal coupling between gravity and the electromagnetic field as well as the self-interaction of the Maxwell field, which can be specifically expressed as
\begin{equation}
	\begin{split}
		\Delta \boldsymbol{\mathcal{L}}_1 = & \frac{1}{16 \pi} \left(\alpha_1 R F_{ab} F^{ab} + \alpha_2 R_{ab} F^{ac} F^{b}_{\ c} + \alpha_3 R_{abcd} F^{ab} F^{cd} \right. \\
		&\left. + \alpha_4 F_{ab} F^{ab} F_{cd} F^{cd} + \alpha_5 F_{ab} F^{bc} F_{cd} F^{da}\right) \boldsymbol{\epsilon}\,,
	\end{split}
\end{equation}
where the parameter $\alpha_i \ \left(i = 1, 2 , 3, 4, 5\right)$ is the coupling constants of two types of the interaction, and the second part of Eq. (\ref{correctterm}) is the quadratic curvature correction which is given as
\begin{equation}
	\Delta \boldsymbol{\mathcal{L}}_{\text{QC}} = \frac{1}{16 \pi} \left( \beta_1 R^2 + \beta_2 R_{ab} R^{ab} + \beta_3 R_{abcd} R^{abcd} \right) \boldsymbol{\epsilon}\,,
\end{equation}
where $\beta_i \ \left(i = 1, 2 , 3 \right)$ represent the coupling constant of the gravitational self-interaction. The last part of Eq. (\ref{totallagrangian}) is the Lagrangian of the additional matter fields in spacetime.

For stationary black holes in the general quadric correction theory of gravity, the entropy of black holes can be described by the Wald entropy. According to the definition of the Wald entropy and the form of the Lagrangian, it can be expressed as
\begin{equation}\label{waldentropy}
	\begin{split}
		S_{\text{W}} = S_{\text{EM}} + S_{\text{QC}} = \frac{1}{4} \int_s d^n y \sqrt{\gamma} & \left(\rho_{\text{EM}} + \rho_{\text{QC}} \right)\,.
	\end{split}
\end{equation}
Eq. (\ref{waldentropy}) shows that the Wald entropy of black holes in the general quadric corrected Einstein-Maxwell gravity can be divided into two parts. The first part $S_{\text{EM}}$ is the entropy of black holes in the Einstein-Maxwell gravity with the non-minimal coupling interaction between gravity and the electromagnetic field as well as the self-interaction of the Maxwell field, the second part $S_{\text{QC}}$ is the entropy that comes from the quadratic curvature correction. While $\rho_{\text{EM}}$ and $\rho_{\text{QC}}$ are densities of the two types of entropy. Since the entropy of black holes in $F (\text{Riemann})$ gravity generally obeys the linearized second law shown in Refs \cite{Chatterjee:2011wj, Wall:2015raa}, the second part of the Wald entropy obeys the linearized second law naturally. To investigate the linearized second law of black holes in the general quadric corrected Einstein-Maxwell gravity, it is not hard to think that we only need to examine whether the first part of the Wald entropy $S_{\text{EM}}$ still satisfies the linearized second law of thermodynamics. In other words, the Lagrangian that is used to examine the linearized second law is given as
\begin{equation}\label{usedlagrangian}
	\boldsymbol{\mathcal{L}} =\boldsymbol{\mathcal{L}}_{\text{EM}} + \Delta \boldsymbol{\mathcal{L}}_1 + \boldsymbol{\mathcal{L}}_{\text{mt}}\,.
\end{equation}

From the Lagrangian of Eq. (\ref{usedlagrangian}), the equation of motion of the gravitational part can be formally written as
\begin{equation}\label{equationofmotion}
		H_{ab} = 8 \pi T_{ab}\,.
\end{equation}
In which the specific expression of $H_{ab}$ is
\begin{equation}\label{exphab}
	H_{ab} = G_{ab} + \frac{1}{2} \sum_{i = 1}^{3} \alpha_i H_{ab}^{(i)}
\end{equation}
with
\begin{equation}\label{hkki}
	\begin{split}
		H_{ab}^{(1)} = & 2R_{ab} F_{cd} F^{cd} - 4 R F_{a}^{\ c} F_{bc} - g_{ab} R F_{cd} F^{cd} - 2 F^{cd} \nabla_a \nabla_b F_{cd} \\
		& - 4 \nabla_a  F^{cd} \nabla_b F_{cd} - 2 F^{cd} \nabla_b \nabla_a F_{cd} + 4 g_{ab} F^{cd} \nabla_e \nabla^e F_{cd} \\
		& + 4 g_{ab} \nabla_e F_{cd} \nabla^e F^{cd}\,, \\
		H_{ab}^{(2)} = & \nabla_a F_b^{\ c} \nabla_d F_c^{\ d} - 2 R_{cd} F_{a}^{\ c} F_{b}^{\ d} -  g_{ab} F^{cd} R_{de} F_{c}^{\ e} +  \nabla_b F_a^{\ c} \nabla_d F_c^{\ d} \\
		& + F^{cd} \nabla_d \nabla_a F_{bc} +  F_b^{\ c} \nabla_d \nabla_a F_c^{\ d} +  F^{cd} \nabla_d \nabla_b F_{ac} + F_a^{\ c} \nabla_d \nabla_b F_c^{\ d} \\
		& + F_b^{\ c} \nabla_d \nabla^d F_{ac} +  F_a^{\ c} \nabla_d \nabla^d F_{bc} +  g_{ab} F^{cd} \nabla_d \nabla_e F_{c}^{\ e} +  \nabla_b F_{cd} \nabla^d F^{\ c}_a \\
		& +  \nabla_a F_{cd} \nabla^d F_b^{\ c} -  g_{ab} \nabla_c F^{cd} \nabla_e F_d^{\ e} +  g_{ab} F^{cd} \nabla_e \nabla_d F_c^{\ e} + g_{ab} \nabla_d F_{ce} \nabla^e F^{cd} \\
		& + 2 \nabla_d F_{bc} \nabla^d F_a^{\ c}\,, \\
		H_{ab}^{(3)} = & 2 F_{b}^{\ c} \nabla_c \nabla_d F_{a}^{\ d} - R_{acde} F_{b}^{\ c} F^{de} - F_{a}^{\ c} F^{de} R_{bcde} - g_{ab} R_{cdef} F^{cd} F^{ef} \\
		& + 2 F_a^{\ c} \nabla_c \nabla_d F_b^{\ d}+ 4 \nabla_c F_a^{\ c} \nabla_d F_{b}^{\ d} + 2 F_b^{\ c} \nabla_d \nabla_c F_a^{\ d} + 2 F_a^{\ c} \nabla_d \nabla_c F_b^{\ d} \\
		& + 4 \nabla_c F_{bd} \nabla^d F_{a}^{\ c}\,,
	\end{split}
\end{equation}
and $G_{ab} = R_{ab} - 1/2 R g_{ab}$ is the Einstein tensor. $T_{ab}$ in Eq. (\ref{equationofmotion}) is the total stress-energy tenor of the theory, it can be decomposed as
\begin{equation}
	T_{ab} = T_{ab}^{\text{EM}} + T_{ab}^{\text{mt}}\,,
\end{equation}
where $T_{ab}^{\text{EM}}$ represents the stress-energy tenor of the electromagnetic field, and $T_{ab}^{\text{mt}}$ is the stress-energy tensor of extra matter fields.

According to Eq. (\ref{usedlagrangian}) and the definition of the Wald entropy, the first part of the Wald entropy $S_{EM}$ is given as
\begin{equation}\label{firstpartofwaldentropy}
	S_{EM} = \frac{1}{4} \int_s d^n y \sqrt{\gamma} \rho_{EM}\,,
\end{equation}
and the concrete expression of the density $\rho_{EM} $ can be written as
\begin{equation}\label{densitywaldentropy}
	\begin{split}
		\rho_{EM} = & 1 + \alpha_1 \left(F^{\hat{a} \hat{b}} F_{\hat{a} \hat{b}} - 2 F_{ac} F_{bd} k^a k^b l^c l^d - 4 F_{a \hat{c}} F_{b}^{\ \hat{c}} k^a l^b \right) \\
		& - \alpha_2 \left(F_{ac} F_{bd} k^a k^b l^c l^d + F_{a \hat{c}} F_{b}^{\ \hat{c}} k^a l^b \right) - 2 \alpha_3 F_{ac} F_{bd} k^a k^b l^c l^d\,,
	\end{split}
\end{equation}
where we have used the symbol `` \ $\hat{}$ \ '' to label the spatial index, i.e., $\gamma_{a}^{\ c} \gamma_{b}^{\ d} X_{cd} = X_{\hat{a} \hat{b}}$, to simplify the expressions. Comparing with the pure Einstein-Maxwell theory, we can see that the non-minimal coupling terms directly affect the expression of the entropy, and it is no longer proportional to the area of the cross section of the event horizon. Therefore, to examine the linearized second law, we only need to investigate whether the expression of Eq. (\ref{firstpartofwaldentropy}) still obeys the linearized second law of black hole thermodynamics under the first-order approximation of the matter fields perturbation.

\section{Examining the linearized second law of black holes in general quadric corrected Einstein-Maxwell gravity}\label{sec3}

Since the non-minimal coupling terms in the Lagrangian of the general quadric corrected Einstein-Maxwell gravity can directly affect the expression of the Wald entropy of black holes, we would like to examine whether the Wald entropy in the corrected gravitational theory still obeys the linearized second law of black hole thermodynamics. To examine the linearized second law, a slow physical accretion process should be introduced first as well. This process is described as extra matter fields pass through the event horizon and fall into the black hole during the quasistationary accretion process. It implies that the process can be regarded as a perturbation for black holes by extra matter fields. Furthermore, it should be required that after the perturbation process, the spacetime geometry of black holes must finally settle down to a stationary state.

In ($\text{n}+2$)-dimensional general quadric corrected Einstein-Maxwell gravity, the event horizon of black holes is denoted as $\mathcal{H}$, which is a ($\text{n}+1$)-dimensional null hypersurface. When a parameter $\lambda$ is chosen as an affine parameter on the event horizon, the event horizon can be generated by the null vector field $k^a = \left(\partial / \partial \lambda \right)^a$, while the null vector field $k^a$ satisfies the geodesic equation $k^b \nabla_b k^a = 0$. { It can be demonstrate that the choice of the null vector field does not influence on the examination of the linearized second law \cite{Kolekar:2012tq, Jiang:2020huc}.} For any cross section on the event horizon, a coordinates with two null fields $\{k^a, l^a, y^a \}$ can be constructed, where $l^a$ is a second null vector, and $y^a$ represents the transverse coordinates on any slice. The relationships between two null vector fields $k^a$ and $l^a$ are
\begin{equation}
	k^a k_a = l^a l_a = 0\,, \qquad k^a l_a = -1\,.
\end{equation}
Based on the two null vectors, the binormal of the cross section is defined as $\epsilon_{ab} = 2 k_{[a} l_{b]}$, and the induced metric on any cross section of the future event horizon is defined as
\begin{equation}
	\gamma_{ab} = g_{ab} + 2 k_{(a} l_{b)}\,.
\end{equation}
Since $\gamma_{ab}$ is the spatial metric, the relationship between the two null vector and the induced metric can be expressed as $k^a \gamma_{ab} = l^a \gamma_{ab} = 0$. For any spatial tensor $X_{a_1 a_2 \cdots}$, using the induced metric $\gamma_{ab}$, one can define the spatial derivative operator $D_a$ as
\begin{equation}
	D_a X_{a_1 a_2 \cdots} = \gamma_{a}^{\ b} \gamma_{a_1 \cdots}^{\ b_1 \cdots} \nabla_b X_{b_1 b_2 \cdots}\,,
\end{equation}
and the spatial derivative operator is compatible to the induce metric, $D_{c} \gamma_{ab} = 0$.

The extrinsic curvature of the event horizon is defined as
\begin{equation}
	B_{ab} = \gamma_{a}^{\ c} \gamma_{b}^{\ d} \nabla_c k_d\,,
\end{equation}
and the evolution of the induced metric along the future event horizon can be obtained as
\begin{equation}
	\gamma_{a}^{\ c} \gamma_{b}^{\ d} \mathcal{L}_k \gamma_{cd} = 2 \left(\sigma_{ab} + \frac{\theta}{n} \gamma_{ab} \right) = 2 B_{ab}\,,
\end{equation}
where $\sigma_{ab}$ and $\theta$ represents the shear and the expansion of the event horizon respectively. The evolution of the extrinsic curvature along the future event horizon can also be obtained as
\begin{equation}\label{evolubab}
	\gamma_{a}^{\ c} \gamma_{b}^{\ d} \mathcal{L}_k B_{cd} = B_{ac} B_{b}^{\ c} - \gamma_{a}^{\ c} \gamma_{b}^{\ d} R_{ecfd} k^e k^f\,.
\end{equation}
From this result, the Raychaudhuri equation is given as
\begin{equation}
	\frac{d \theta}{d \lambda} = - \frac{\theta^2}{n - 2} - \sigma_{ab} \sigma^{ab} - R_{kk}\,,
\end{equation}
where we have used the convention $X_{kk} = X_{ab} k^a k^b$ for any tensor $X_{ab}$.

To represent the perturbation that comes from black holes absorbing matter fields, the sufficient small parameter $\epsilon$ is introduced to describe the order of the approximation of the perturbation. According to the small parameter, we assume that $B_{ab} \sim \theta \sim \sigma_{ab} \sim \mathcal{O} \left(\epsilon \right)$. Since we only would like to consider the linearized second law during the dynamical process, the symbol ``$\simeq$'' will be used to represent the identity under the linear order approximation of the matter fields perturbation in the following calculation.

Considering the first-order approximation of the dynamical perturbation, the linear version of the Raychaudhuri equation can be written as
\begin{equation}
	\frac{d \theta}{d \lambda} \simeq - R_{kk}\,,
\end{equation}
and Eq. (\ref{evolubab}) is simplified as
\begin{equation}
	\gamma_{a}^{\ c} \gamma_{b}^{\ d} \mathcal{L}_k B_{cd} \simeq - \gamma_{a}^{\ c} \gamma_{b}^{\ d} R_{ecfd} k^e k^f\,.
\end{equation}

From the physical perspective, the total stress-energy tensor should be satisfied the null energy condition. Considering any null vector field $n^a$, the null energy condition of $T_{ab}$ can be expressed as $T_{ab} n^a n^b \geq 0$. It indicates that two parts of the total stress-energy tensor are both obeying the null energy condition. An assumption should be added in our demonstration that the matter fields always satisfy the null energy condition. According to the calculation in Appendix \ref{appa}, it is shown that in the coordinates with two null vectors $k^a$ and $l^a$, the total stress-energy tensor under the first-order approximation of the perturbation can be expressed as
\begin{equation}
	T_{ab} k^a k^b \simeq T_{ab}^{\text{mt}} k^a k^b \ge 0\,.
\end{equation}
It implies that when we only consider the linear order approximation, the total stress-energy tensor obeys the requirement of the null energy condition naturally.

From now on, we would like to examine whether the first part of the Wald entropy still satisfies the linearized second law during the matter fields perturbation. Following a similar train of thought of Ref. \cite{Kolekar:2012tq}, when the linearized second law is valid during the perturbation process while the total stress-energy tensor obeys the null energy condition, the entropy of black holes should be satisfied the following relationship under the first-order approximation, i.e.,
\begin{equation}
	\begin{split}
		\mathcal{L}_k^2 S \simeq - 2 \pi \int_s \tilde{\epsilon} T_{kk} = - \frac{1}{4} \int_s \tilde{\epsilon} H_{kk} < 0\,.
	\end{split}
\end{equation}
Since we have assumed that the evolution of black holes eventually tends to a stationary state after the dynamical process, it implies that the rate of change of the entropy should vanish in the asymptotic future. In other words, the second-order Lie derivative of the entropy is negative, $\mathcal{L}_k^2 S < 0 $, during the process. Meanwhile, the decrease of the rate of change also implies that the value of the entropy of black holes always increases with the perturbation process in the future null direction, i.e., $\mathcal{L}_k S > 0$. To investigate whether the first part of the Wald entropy obeys the linearized second law, we only need to examine whether the first part of the Wald entropy satisfies the relationship
\begin{equation}
	\mathcal{L}_k^2 S_{\text{EM}} \simeq - \frac{1}{4} \int_s \tilde{\epsilon} H_{kk} < 0
\end{equation}
under the linear order approximation. Therefore, in the following, we need to obtain the expression of the integral of $H_{kk}$ on an arbitrary cross section of the event horizon under the linear order approximation.

From Eq. (\ref{exphab}), the specific expression of $H_{kk}$ is given as
\begin{equation}\label{exptotalhkk}
	\begin{split}
		H_{kk} = & R_{kk} + \frac{1}{2} \sum_{i = 1}^3 \alpha_i H_{kk}^{(i)}\,,
	\end{split}
\end{equation}
where
\begin{equation}\label{ohkk1}
	\begin{split}
		H_{kk}^{(1)} = & 2 R_{kk} F_{cd} F^{cd} - 4 R k^a k^b F_{a}^{\ c} F_{bc} - 2 k^a k^b \nabla_a \nabla_b \left(F^{cd} F_{cd} \right)\,,
	\end{split}
\end{equation}
\begin{equation}\label{ohkk2}
	\begin{split}
		H_{kk}^{(2)} = & k^a k^b \nabla_a F_{b}^{\ c} \nabla_d F_{c}^{\ d} - 2 k^a k^b F_{a}^{\ c} F_b^{\ d} R_{cd} + k^a k^b \nabla_b F_{a}^{\ c} \nabla_d F_{c}^{\ d} \\
		& + k^a k^b F^{cd} \nabla_d \nabla_a F_{bc} + k^a k^b F_{b}^{\ c} \nabla_d \nabla_a F_{c}^{\ d} + k^a k^b F^{cd} \nabla_d \nabla_b F_{ac} \\
		& + k^a k^b F_{a}^{\ c} \nabla_d \nabla_b F_{c}^{\ d} + k^a k^b F_{b}^{\ c} \nabla_d \nabla^d F_{ac} + k^a k^b F_{a}^{\ c} \nabla_d \nabla^d F_{bc}\\
		& + k^a k^b \nabla_b F_{cd} \nabla^d F_{a}^{\ c} + k^a k^b \nabla_a F_{cd} \nabla^d F_{b}^{\ c} + 2 k^a k^b \nabla_d F_{bc} \nabla^d F_{a}^{\ c}\,,
	\end{split}
\end{equation}
\begin{equation}\label{ohkk3}
	\begin{split}
		H_{kk}^{(3)} = & 2 k^a k^b F_{b}^{\ c} \nabla_c \nabla_d F_{a}^{\ d} - k^a k^b R_{acde} F_{b}^{\ c} F^{de} - k^a k^b R_{bcde} F_{a}^{\ c} F^{de} \\
		& + 2 k^a k^b F_{a}^{\ c} \nabla_c \nabla_d F_{b}^{\ d} + 4 k^a k^b \nabla_c F_{a}^{\ c} \nabla_d F_{b}^{\ d} + 2 k^a k^b F_{b}^{\ c} \nabla_d \nabla_c F_{a}^{\ d}\\
		& +2 k^a k^b F_{a} ^{\ c} \nabla_d \nabla_c F_{b}^{\ d} + 4 k^a k^b \nabla_c F_{bd} \nabla^d F_{a}^{\ c}\,.
	\end{split}
\end{equation}
For the integral form on any slice of the first term in Eq. (\ref{exptotalhkk}), the integrand under the first-order approximation is given as
\begin{equation}\label{rkksqgamma}
	\begin{split}
		R_{kk} \sqrt{\gamma} = & - \left(\mathcal{L}_k \theta \right) \sqrt{\gamma} \\
		= & - \mathcal{L}_k \left(\theta \sqrt{\gamma} \right) + \theta \left(\mathcal{L}_k \sqrt{\gamma} \right)\\
		= & - \mathcal{L}_k^2 \sqrt{\gamma} + \theta^2 \sqrt{\gamma}\\
		\simeq & - \mathcal{L}_k^2 \sqrt{\gamma}\,.
	\end{split}
\end{equation}
The integral form of the first term in Eq. (\ref{exptotalhkk}) can be expressed as
\begin{equation}\label{finalresultrkk}
	\begin{split}
		\int_s \tilde{\epsilon} R_{kk} \simeq \mathcal{L}_k^2 \int_s \tilde{\epsilon} \left(- 1\right)\,.
	\end{split}
\end{equation}

According to Appendix \ref{appb}, after retaining the linear order terms as well as neglecting the high-order terms, the $H_{kk}^{(1)}$ can be finally written as
\begin{equation}\label{hkk1result}
	\begin{split}
		H_{kk}^{(1)} = 2 R_{kk} F_{cd} F^{cd} - 2 \mathcal{L}_k^2 \left(F^{cd} F_{cd} \right)\,.
	\end{split}
\end{equation}
To obtain the integral form of $H_{kk}^{(1)}$ on the cross-section, the integrand should be simplified firstly. Combining with the induced volume element and using the result of Eq. (\ref{rkksqgamma}), the integrand in the integral form of $H_{kk}^{(1)}$ can be calculated as
\begin{equation}
	\begin{split}
		H_{kk}^{(1)} \sqrt{\gamma} = & 2 R_{kk} F_{cd} F^{cd} \sqrt{\gamma} - 2 \mathcal{L}_k^2 \left(F^{cd} F_{cd} \right) \sqrt{\gamma} \\
		\simeq & - 2 F_{cd} F^{cd} \left( \mathcal{L}_k^2 \sqrt{\gamma} \right) - 2 \mathcal{L}_k^2 \left(F^{cd} F_{cd} \right) \sqrt{\gamma}\\
		\simeq & - 2 \mathcal{L}_k^2 \left[ \left(F_{cd} F^{cd} \right) \sqrt{\gamma} \right] + 2 \mathcal{L}_k^2 \left(F_{cd} F^{cd} \right) \sqrt{\gamma} \\
		& + 4 \mathcal{L}_k \left(F_{cd} F^{cd} \right) \left(\mathcal{L}_k \sqrt{\gamma} \right) - 2 \mathcal{L}_k^2 \left(F_{cd} F^{cd} \right) \sqrt{\gamma}\\
		\simeq & \mathcal{L}_k^2 \left[- 2 \left(F_{cd} F^{cd} \right) \sqrt{\gamma} \right]\,.
	\end{split}
\end{equation}
Expanding duplicated indexes in the tensors of electromagnetic field, the integral form of $H_{kk}^{(1)}$ can be finally written as
\begin{equation}\label{finalresulthkk1}
	\begin{split}
		\int_s \tilde{\epsilon} H_{kk}^{(1)} \simeq \mathcal{L}_k^2 \int_s \tilde{\epsilon} \left[4 \left(k^a F_{ae} l^e \right) \left(k^b F_{bf} l^f \right) - 2 F_{\hat{c} \hat{d}} F^{\hat{c} \hat{d}} + 8 \left(k^a F_{a \hat{e}} \right) \left(l^d F_{d}^{\ \hat{e}} \right) \right]\,.
	\end{split}
\end{equation}

According to Appendix \ref{appc}, the expression of $H_{kk}^{(2)}$ under the first-order approximation of the dynamical process is given as
\begin{equation}\label{firordhkk2}
	\begin{split}
		H_{kk}^{(2)} \simeq & - 2 \left[\mathcal{L}_k \left(k^b F_{b}^{\ \hat{e}} \right) \right] \left[D_{\hat{e}} \left(k^a F_{ad} l^d \right) \right]  - 2 \left(k^a F_{ad} l^d \right) D^{\hat{e}} \left[\mathcal{L}_k \left(k^b F_{b\hat{e}} \right) \right] \\
		& + 2 \mathcal{L}_k \left(k^a F_{a\hat{c}} \right) D_{\hat{e}} \left( F^{\hat{c} \hat{e}} \right) + 2 \left(F^{\hat{e} \hat{d}} \right) D_{\hat{d}} \left[\mathcal{L}_k \left(k^a F_{a \hat{e}} \right) \right] \\
		& - 4 \left(k^a F_{a}^{\ \hat{e}} \right) \left[ \mathcal{L}_k^2 \left(F_{\hat{e}d} l^d \right) \right] + 4 \left(k^a F_{a}^{\ \hat{e}} \right) \left[\mathcal{L}_k^2 \left(F_{\hat{e}d} l^d \right) \right] \\
		& - 2 \left(k^a F_{ae} l^e \right) \left(k^b F_{bf} l^f \right) \left(R_{kk} \right) + 4 \left(k^a F_{ae} l^e \right) \left[\mathcal{L}_k^2 \left(k^b F_{bf} l^f \right) \right] \\
		& + 2 \left(l^d F_{d}^{\ \hat{e}} \right) \left[\mathcal{L}_k^2 \left(k^a F_{a \hat{e}} \right) \right] + 2 \left(k^a F_{a \hat{e}} \right) \left[\mathcal{L}_k^2 \left(l^d F_{d}^{\ \hat{e}} \right) \right] \\
		& + 4 \left[\mathcal{L}_k \left(k^a F_{a \hat{e}} \right) \right] \left[\mathcal{L}_k \left(l^d F_{d}^{\ \hat{e}} \right) \right]\,.
	\end{split}
\end{equation}
The first four terms of Eq. (\ref{firordhkk2}) can be simplified as
\begin{equation}
	\begin{split}
		& - 2 \left[\mathcal{L}_k \left(k^b F_{b}^{\ \hat{e}} \right) \right] \left[D_{\hat{e}} \left(k^a F_{ad} l^d \right) \right]  - 2 \left(k^a F_{ad} l^d \right) D^{\hat{e}} \left[\mathcal{L}_k \left(k^b F_{b\hat{e}} \right) \right] \\
		& + 2 \mathcal{L}_k \left(k^a F_{a\hat{c}} \right) D_{\hat{e}} \left( F^{\hat{c} \hat{e}} \right) + 2 \left(F^{\hat{e} \hat{d}} \right) D_{\hat{d}} \left[\mathcal{L}_k \left(k^a F_{a \hat{e}} \right) \right] \\
		= & - 2 D_{\hat{e}} \left[\left(k^a F_{ad} l^d \right) \mathcal{L}_k \left(k^b F_{b}^{\ \hat{e}} \right)  \right] + 2 \left(k^a F_{ad} l^d \right) D^{\hat{e}} \left[\mathcal{L}_k \left(k^b F_{b\hat{e}} \right) \right]\\
		& - 2 \left(k^a F_{ad} l^d \right) D^{\hat{e}} \left[\mathcal{L}_k \left(k^b F_{b\hat{e}} \right) \right] + 2 D_{\hat{e}} \left[\left( F^{\hat{c} \hat{e}} \right) \mathcal{L}_k \left(k^a F_{a\hat{c}} \right) \right] \\
		& - 2 \left(F^{\hat{c} \hat{e}} \right) D_{\hat{e}} \left[\mathcal{L}_k \left(k^a F_{a \hat{c}} \right) \right] + 2 \left(F^{\hat{c} \hat{e}} \right) D_{\hat{e}} \left[\mathcal{L}_k \left(k^a F_{a \hat{c}} \right) \right]\\
		= & - 2 D_{\hat{e}} \left[\left(k^a F_{ad} l^d \right) \mathcal{L}_k \left(k^b F_{b}^{\ \hat{e}} \right)  \right] + 2 D_{\hat{e}} \left[\left( F^{\hat{c} \hat{e}} \right) \mathcal{L}_k \left(k^a F_{a\hat{c}} \right) \right]\,.
	\end{split}
\end{equation}
Based on the assumption that the event horizon is compact, the surface term in the integral on the cross-section of the event horizon can be ignored directly. It means that the first four terms of Eq. (\ref{firordhkk2}) do not contributed to the final result. The fifth and sixth terms of Eq. (\ref{firordhkk2}) can be canceled with each other. Utilizing the result of Eq. (\ref{rkksqgamma}) again, the seventh and eighth terms of Eq. (\ref{firordhkk2}) with the induced volume element can be calculated as
\begin{equation}
	\begin{split}
		& - 2 \left(k^a F_{ae} l^e \right) \left(k^b F_{bf} l^f \right) \left(R_{kk} \right) \sqrt{\gamma} + 4 \left(k^a F_{ae} l^e \right) \mathcal{L}_k^2 \left(k^b F_{bf} l^f \right) \sqrt{\gamma} \\
		\simeq & 2 \left(k^a F_{ae} l^e \right) \left(k^b F_{bf} l^f \right) \mathcal{L}_k^2 \sqrt{\gamma} + 2 \mathcal{L}_k^2 \left[ \left(k^a F_{ae} l^e \right) \left(k^b F_{bf} l^f \right)  \right] \sqrt{\gamma}\\
		\simeq & 2 \mathcal{L}_k^2 \left[\left(k^a F_{ae} l^e \right) \left(k^b F_{bf} l^f \right) \sqrt{\gamma} \right] - 2 \mathcal{L}_k^2 \left[\left(k^a F_{ae} l^e \right) \left(k^b F_{bf} l^f \right)\right]  \sqrt{\gamma} \\
		& - 4  \mathcal{L}_k \left[\left(k^a F_{ae} l^e \right) \left(k^b F_{bf} l^f \right)\right] \theta \sqrt{\gamma} + 2 \mathcal{L}_k^2 \left[\left(k^a F_{ae} l^e \right) \left(k^b F_{bf} l^f \right)\right]  \sqrt{\gamma}\\
		\simeq &  2 \mathcal{L}_k^2 \left[\left(k^a F_{ae} l^e \right) \left(k^b F_{bf} l^f \right) \sqrt{\gamma} \right]\,.
	\end{split}
\end{equation}
Therefore, the integral form of the seventh and eighth terms in Eq. (\ref{firordhkk2}) is obtained as
\begin{equation}
	\begin{split}
		& \int_s  \tilde{\epsilon} \left[- 2 \left(k^a F_{ae} l^e \right) \left(k^b F_{bf} l^f \right) \left(R_{kk} \right) + 4 \left(k^a F_{ae} l^e \right) \mathcal{L}_k^2 \left(k^b F_{bf} l^f \right)  \right]\\
		= & \mathcal{L}_k^2 \int_s \tilde{\epsilon} \left[2 \left(k^a F_{ae} l^e \right) \left(k^b F_{bf} l^f \right) \right]\,.
	\end{split}
\end{equation}
For the last three terms of Eq. (\ref{firordhkk2}), the integrand of these terms under the first-order approximation can be simplified as
\begin{equation}
	\begin{split}
		& 2 \left(l^d F_{d}^{\ \hat{e}} \right) \left[\mathcal{L}_k^2 \left(k^a F_{a \hat{e}} \right) \right] \sqrt{\gamma} + 2 \left(k^a F_{a \hat{e}} \right) \left[\mathcal{L}_k^2 \left(l^d F_{d}^{\ \hat{e}} \right) \right] \sqrt{\gamma} \\
		& + 4 \left[\mathcal{L}_k \left(k^a F_{a \hat{e}} \right) \right] \left[\mathcal{L}_k \left(l^d F_{d}^{\ \hat{e}} \right) \right]\sqrt{\gamma}\\
		= & 2 \mathcal{L}_k^2 \left[\left(k^a F_{a \hat{e}} \right) \left(l^d F_{d}^{\ \hat{e}} \right) \right] \sqrt{\gamma}\\
		= & 2 \mathcal{L}_k^2 \left[\left(k^a F_{a \hat{e}} \right) \left(l^d F_{d}^{\ \hat{e}} \right) \sqrt{\gamma} \right] - 4  \mathcal{L}_k \left[\left(k^a F_{a \hat{e}} \right) \left(l^d F_{d}^{\ \hat{e}} \right) \right] \theta \sqrt{\gamma} \\
		& - 2 \left[\left(k^a F_{a \hat{e}} \right) \left(l^d F_{d}^{\ \hat{e}} \right) \right] \left(\mathcal{L}_k \theta \right) \sqrt{\gamma} - 2 \left[\left(k^a F_{a \hat{e}} \right) \left(l^d F_{d}^{\ \hat{e}} \right) \right] \theta^2 \sqrt{\gamma} \\
		\simeq & 2 \mathcal{L}_k^2 \left[\left(k^a F_{a \hat{e}} \right) \left(l^d F_{d}^{\ \hat{e}} \right) \sqrt{\gamma} \right]\,.
	\end{split}
\end{equation}
The integral form of the last three terms can be further written as
\begin{equation}
	\begin{split}
		& \int_s \tilde{\epsilon} \left[2 \left(l^d F_{d}^{\ \hat{e}} \right) \mathcal{L}_k^2 \left(k^a F_{a \hat{e}} \right)  + 2 \left(k^a F_{a \hat{e}} \right) \mathcal{L}_k^2 \left(l^d F_{d}^{\ \hat{e}} \right) + 4 \mathcal{L}_k \left(k^a F_{a \hat{e}} \right) \mathcal{L}_k \left(l^d F_{d}^{\ \hat{e}} \right) \right]\\
		= & \mathcal{L}_k^2 \int_s \tilde{\epsilon} \left[2 \left(k^a F_{a \hat{e}} \right) \left(l^d F_{d}^{\ \hat{e}} \right) \right]\,.
	\end{split}
\end{equation}
Therefore, the integral form of $H_{kk}^{(2)}$ can be finally expressed as
\begin{equation}\label{finalresulthkk2}
	\begin{split}
		\int_s \tilde{\epsilon} H_{kk}^{(2)} =  \mathcal{L}_k^2  \int_s \tilde{\epsilon} \left[2 \left(k^a F_{ae} l^e \right) \left(k^b F_{bf} l^f \right) + 2 \left(k^a F_{a \hat{e}} \right) \left(l^d F_{d}^{\ \hat{e}} \right) \right]\,.
	\end{split}
\end{equation}

The result of $H_{kk}^{(3)}$ under the first-order approximation is given by Appendix \ref{appd}, the specific expression can be expressed as
\begin{equation}\label{firstorderhkk3}
	\begin{split}
		H_{kk}^{(3)} \simeq & - 8 \left(k^a F_{ad} l^d \right) D^{\hat{e}} \left[\mathcal{L}_k \left(k^b F_{b \hat{e}} \right) \right] - 8 \mathcal{L}_k \left(k^b F_{b \hat{c}} \right) D^{\hat{c}} \left(k^a F_{ad} l^d \right) \\
		& - 4 \left(k^c F_{ce} l^e \right) \left(k^d F_{df} l^f \right) \left(R_{kk} \right) + 8 \left(k^a F_{ae} l^e \right) \left[\mathcal{L}_k^2 \left(k^b F_{bf} l^f \right) \right]\,.
	\end{split}
\end{equation}
According to the compactness of the event horizon, the integral form of the first two terms in Eq. (\ref{firstorderhkk3}) can be simplified as
\begin{equation}
	\begin{split}
		& \int_s \tilde{\epsilon} \left[- 8 \left(k^a F_{ad} l^d \right) D^{\hat{e}} \left[\mathcal{L}_k \left(k^b F_{b \hat{e}} \right) \right] - 8 \mathcal{L}_k \left(k^b F_{b \hat{c}} \right) D^{\hat{c}} \left(k^a F_{ad} l^d \right) \right]\\
		= &- 8 \int_s \tilde{\epsilon}  D^{\hat{c}} \left[ \left(k^a F_{ad} l^d \right) \mathcal{L}_k \left(k^b F_{b \hat{c}} \right) \right] + 8 \int_s \tilde{\epsilon} D^{\hat{c}} \left(k^a F_{ad} l^d \right) \mathcal{L}_k \left(k^b F_{b \hat{c}} \right) \\
		& - 8  \int_s \tilde{\epsilon} \mathcal{L}_k \left(k^b F_{b \hat{c}} \right) D^{\hat{c}} \left(k^a F_{ad} l^d \right)\\
		= & 0\,,
	\end{split}
\end{equation}
The integral of $H_{kk}^{(3)}$ is further given as
\begin{equation}\label{hkk3finalexpression}
	\begin{split}
		\int_s \tilde{\epsilon} H_{kk}^{(3)} = & \int_s  \tilde{\epsilon} \left[- 4 \left(k^c F_{ce} l^e \right) \left(k^d F_{df} l^f \right) \left(R_{kk} \right) + 8 \left(k^a F_{ae} l^e \right) \left[\mathcal{L}_k^2 \left(k^b F_{bf} l^f \right)  \right] \right]\,.
	\end{split}
\end{equation}
Using the result of Eq. (\ref{rkksqgamma}), the integrand in Eq. (\ref{hkk3finalexpression}) can be simplified as
\begin{equation}
	\begin{split}
		H_{kk}^{(3)} \sqrt{\gamma} = & - 4 \left(k^c F_{ce} l^e \right) \left(k^d F_{df} l^f \right) \left(R_{kk} \right) \sqrt{\gamma} + 8 \left(k^a F_{ae} l^e \right) \left[\mathcal{L}_k^2 \left(k^b F_{bf} l^f \right)  \right] \sqrt{\gamma} \\
		\simeq & 4 \left(k^a F_{ab} l^b \right) \left(k^c F_{cd} l^d \right) \mathcal{L}_k^2 \sqrt{\gamma} + 4 \mathcal{L}_k^2 \left[\left(k^a F_{ab} l^b \right) \left(k^c F_{cd} l^d \right) \right] \sqrt{\gamma} \\
		\simeq & 4 \mathcal{L}_k^2 \left[\left(k^a F_{ab} l^b \right) \left(k^c F_{cd} l^d \right) \sqrt{\gamma} \right] - 4 \mathcal{L}_k^2  \left[\left(k^a F_{ab} l^b \right) \left(k^c F_{cd} l^d \right) \right] \sqrt{\gamma} \\
		& - 4 \mathcal{L}_k \left[\left(k^a F_{ab} l^b \right) \left(k^c F_{cd} l^d \right) \right] \mathcal{L}_k \sqrt{\gamma} + 4 \mathcal{L}_k^2  \left[\left(k^a F_{ab} l^b \right) \left(k^c F_{cd} l^d \right) \right] \sqrt{\gamma} \\
		\simeq & \mathcal{L}_k^2 \left[4 \left(k^a F_{ab} l^b \right) \left(k^c F_{cd} l^d \right) \sqrt{\gamma} \right]\,.
	\end{split}
\end{equation}
Therefore, the integral form of $H_{kk}^{(3)}$ can be finally obtained as
\begin{equation}
	\begin{split}\label{finalresulthkk3}
		\int_s  \tilde{\epsilon} H_{kk}^{(3)} = \mathcal{L}_k^2 \int_s  \tilde{\epsilon} \left[4 \left(k^a F_{ab} l^b \right) \left(k^c F_{cd} l^d \right) \right]\,.
	\end{split}
\end{equation}
Combining with Eq. (\ref{finalresultrkk}), Eq. (\ref{finalresulthkk1}), Eq. (\ref{finalresulthkk2}), and Eq. (\ref{finalresulthkk3}), the integral form of $H_{kk}$ can be given as
\begin{equation}\label{finalintegralhkk}
	\begin{split}
		\int_s \tilde{\epsilon} H_{kk} = & \int_s \tilde{\epsilon} \left(R_{kk} + \frac{1}{2} \sum_{i = 1}^{3} \alpha_i H_{kk}^{(i)} \right)\\
		\simeq & - \mathcal{L}_k^2 \int_s \tilde{\epsilon} \left[1+ \alpha_1 F_{\hat{c} \hat{d}} F^{\hat{c} \hat{d}} - 2 \alpha_1 \left(k^a F_{ae} l^e \right) \left(k^b F_{bf} l^f \right) - 4 \alpha_1 \left(k^a F_{a \hat{e}} \right) \left(l^d F_{d}^{\ \hat{e}} \right) \right.\\
		& \left. - \alpha_2 \left(k^a F_{ae} l^e \right) \left(k^b F_{bf} l^f \right) - \alpha_2 \left(k^a F_{a \hat{e}} \right) \left(l^d F_{d}^{\ \hat{e}} \right) - 2 \alpha_3 \left(k^a F_{ab} l^b \right) \left(k^c F_{cd} l^d \right) \right]\\
		= & - \mathcal{L}_k^2 \int_s \tilde{\epsilon} \ \rho_{\text{EM}}\,,
	\end{split}
\end{equation}
where we have used Eq. (\ref{densitywaldentropy}) in the last step. After adding coefficient $1 / 4$ to both sides of Eq. (\ref{finalintegralhkk}), we have
\begin{equation}
	\mathcal{L}_k^2 S_{\text{EM}} = - \frac{1}{4} \int_s \tilde{\epsilon} H_{kk} < 0\,.
\end{equation}
This result implies that the value of $S_{\text{EM}}$ always increases, while its growth rate gradually decreases until the end of the perturbation process. In other words, the value of the first part of the Wald entropy increases with the dynamical process, and black holes finally settle down to a stationary state after the perturbation. Since the second part of the Wald entropy $S_{\text{QC}}$ has been demonstrated that it always obeys the linearized second law, the Wald entropy of black holes in general quadric corrected Einstein-Maxwell gravity still satisfies the linearized second law under the first-order approximation of the matter fields perturbation.

\section{Discussions and Conclusions}\label{sec4}
{ Although the Wald entropy of black holes in general quadric corrected Einstein-Maxwell gravity has been generally demonstrated that it always satisfies the linearized second law, there are still some key issues that should be discussed in further detail. In our demonstration, In our demonstration, the Wald entropy has been used to represent the entropy of black holes in the corrected gravitational theory. However, the Wald entropy is only suitable for describing the entropy of black holes at the initial and final states because the two states are both stationary, while the expression of the Wald entropy is independent of the selection of the cross-section on the event horizon. Since the states of black holes are no longer stationary during the perturbation process, the cross-section is evolving with the process all the time. And the expression of the Wald entropy also changes with the evolution of the cross-section according to the definition. This situation makes it impossible to distinguish which expression of the Wald entropy corresponds to the entropy of black holes during the perturbation. The definition of the cross-section comes from the choice of the null generator $k^a$. It means that the evolution of the cross-section is caused by the change of $k^a$ with the perturbation. Fortunately, the proof of the linearized second law is independent of the choice of the null generator as mentioned above. It indicates that our demonstration is valid for any cross-section on the event horizon whether or not the area of the cross-section corresponds to the entropy of black holes. Besides, for the issue of which area of cross-section corresponds to the entropy of black holes during the perturbation process, we cannot solve it only relying on the first and the second laws of thermodynamics unless involving some other constraint conditions.}

The entropy of black holes in any diffeomorphism invariant theory of gravity is generally given by the Wald entropy, while it has been shown that the Wald entropy always obeys the first law of black hole thermodynamics. However, whether the Wald entropy obeys the second law of thermodynamics still requires further study. When we consider a gravitational theory that contains the non-minimal coupling interaction between gravity and matter fields, the expression of the Wald entropy can be sufficiently affected by the non-minimal coupling terms in the Lagrangian. Therefore, examining whether the Wald entropy of black holes in the gravitational theory with the non-minimum coupling matter fields satisfies the second law is an important prerequisite for examining whether it generally satisfies the second law in any diffeomorphism invariant gravity. The general quadric corrected Einstein-Maxwell gravity mainly includes three types of correction terms, which describe self-interaction of the electromagnetic field, the non-minimal coupling interaction between gravity and the Maxwell field, and the quadratic interaction of gravity. According to the definition of the Wald entropy, the non-minimal coupling terms and the quadratic curvature terms will sufficiently affect the form of the Wald entropy, while the expression of the Wald entropy can be divided into the two parts. The first part is the entropy of black holes in the Einstein-Maxwell gravity with non-minimal coupling interaction and the self-interaction of the electromagnetic field, the second part of the Wald entropy comes from the quadratic curvature correction. Since we only want to investigate the linearized second law, the two parts of the Wald entropy can be examined respectively whether it obeys the linearized second law. However, it has shown that the second part of the Wald entropy always obeys the linearized second law. So to investigate the linearized second law of black holes in the general quadric corrected Einstein-Maxwell gravity, we only need to examine whether the first part of the Wald entropy still obeys the linearized second law. To study this issue, a quasistationary accreting process that matter fields pass through the event horizon and perturb the geometry of the black hole is considered, while matter fields are required to satisfy the null energy condition. Two assumptions that black holes will settle down to a stationary state after the perturbation process and that there is a regular bifurcation surface in the background spacetime are further introduced. According to the Raychaudhuri equation, we demonstrate that the second-order Lie derivative of the first part of the Wald entropy is negative under the linear order approximation of the dynamical accreting process. It implies that the value of the entropy must increase along the direction of the future event horizon during the process, while the magnitude of its increase gradually decreases. The result indicates that black holes can eventually tend to a stationary state, and the first part of the Wald entropy always satisfies the requirement of the linearized second law during the perturbation process. Since the two parts of the Wald entropy are obeying the linearized second law, the Wald entropy of black holes in the general quadric correction Einstein-Maxwell gravity always satisfies the linearized second law.

\section*{Acknowledgement}
X.-Y. W. is supported by the Talents Introduction Foundation of Beijing Normal University with Grant No. 111032109. J. J. is supported by the National Natural Science Foundation of China (NSFC) with Grants No. 11775022 and 11873044.

\appendix

\section{Examining the null energy condition of the total stress-energy tensor}\label{appa}
We would like to demonstrate that the total stress-energy tensor in Eq. (\ref{equationofmotion}) should satisfy the null energy tensor in this appendix. The stress energy tensor can be decomposed into the two parts, which can be formally written as
\begin{equation}
	T_{ab} = T_{ab}^{\text{EM}} + T_{ab}^{\text{mt}}\,.
\end{equation}
We have assumed that additional matter fields should always satisfy the null energy condition. It means that for any null vector $n^a$, the null energy condition of the stress-energy tensor of matter fields can be expressed as
\begin{equation}
	T_{ab}^{\text{mt}} n^a n^b \ge 0\,.
\end{equation}
Hence, to prove the total stress-energy tensor $T_{ab}$ obeying the null energy condition, we only need to demonstrate the electromagnetic field part of stress-energy tensor $T_{ab}^{\text{EM}}$ satisfies the null energy condition. From the equation of motion, the specific expression of $T_{ab}^{\text{EM}}$ is given by
\begin{equation}
	\begin{split}
		T_{ab}^{\text{EM}} = & \alpha_4 \left(4 F_{a}^{\ c} F_{bc} F_{de} F^{de} - \frac{1}{2} g_{ab} F_{cd} F^{cd} F_{ef} F^{ef} \right) \\
		& + \alpha_5 \left(4 F_{a}^{\ c} F_{b}^{\ d} F_{c}^{\ e} F_{de} - \frac{1}{2} g_{ab} F_{c}^{\ e} F^{cd} F_{d}^{\ f} F_{ef} \right)\,.
	\end{split}
\end{equation}
According to the coordinates defined in Sec. \ref{sec3}, the electromagnetic field part of stress-energy tensor $T_{ab}^{\text{EM}}$ contracting with two null vectors $k^a$ can be further expressed as
\begin{equation}\label{appenatkk}
	T_{kk}^{\text{EM}} = k^a k^b T_{ab}^{\text{EM}} = 4 \alpha_4 k^a k^b F_{a}^{\ c} F_{bc} F_{de} F^{de} + 4 \alpha_5 F_{a}^{\ c} F_{b}^{\ d} F_{c}^{\ e} F_{de}\,.
\end{equation}

Before investigating the null energy condition of $T_{ab}^{\text{EM}}$, an convention should be introduced first. In the following calculation, we will use the parenthesis or the square brackets with a subscript to represents the order of the expression, i.e., $ ( \ )_n$ or $ [ \ ]_n$, $n = 0 , 1$. The expression is a zero-order or a background quantity when $n = 0$, and the expression is a first-order quantity when $n = 1$. According to the convention to represent the order of the expression, we further examine whether the stress-energy tensor $T_{ab}^{\text{EM}}$ satisfies the null energy condition. Using the definition of the induced metric $\gamma^{ab}$ to expand the index $c$ in the expression, the first term of Eq. (\ref{appenatkk}) can be written as
\begin{equation}
	\begin{split}
		4 \alpha_4 k^a k^b F_{a}^{\ c} F_{bc} F_{de} F^{de} = 4 \alpha_4 \left(k^a F_{a}^{\ \hat{c}} \right)_1 \left(k^b F_{b \hat{c}} \right)_1 \left(F_{ef} F^{ef} \right)_0 \simeq 0\,.
	\end{split}
\end{equation}
Since the result is the multiplication of two first-order terms, the first term of Eq. (\ref{appenatkk}) is the second-order term. The first term of Eq. (\ref{appenatkk}) should be dropped out because only the first-order approximation of the matter fields perturbation is considered in our demonstration. Similarly, the second term in Eq. (\ref{appenatkk}) can be calculated as
\begin{equation}
	\begin{split}
		& 4 \alpha_5 F_{a}^{\ c} F_{b}^{\ d} F_{c}^{\ e} F_{de} \\
		= & 4 \alpha_5 \left(k^a F_{a \hat{c}} \right)_1 \left(k^b F_{b \hat{d}} \right)_1 F^{\hat{c} \hat{e}} F^{\hat{d}}_{\ \hat{e}} - 4 \alpha_5 \left(k^a F_{ae} l^e \right)_0 \left(k^b F_{bf} l^f \right)_0 \left(k^c F_{c \hat{g}} \right)_1 \left(k^d F_{d}^{\ \hat{g}} \right)_1 \\
		& - 8 \alpha_5 \left(k^a F_{a \hat{e}} \right)_1 \left(k^b F_{b}^{\ \hat{e}} \right)_1 \left(k^c F_{c \hat{f}} \right)_1 \left(l^d F_{d}^{\ \hat{f}} \right)_0\\
		\simeq & 0\,.
 	\end{split}
\end{equation}
We can easily see that at least two first-order terms exist in each term of the above equation. It means that the second term in Eq. (\ref{appenatkk}) is also a second-order term. Since the first and the second terms of Eq. (\ref{appenatkk}) are both proportional to the second-order approximation, the electromagnetic part of the stress-energy tensor can be neglected directly when only considering the first-order approximation of the matter fields perturbation. Therefore, utilizing the null energy condition of extra matter fields, the null energy condition of the total stress-energy tensor can be simplified expressed as
\begin{equation}
	T_{kk} \simeq T_{kk}^{\text{mt}} \ge 0
\end{equation}
under the first-order approximation. It is shown that the total stress-energy tensor always satisfies the null energy condition.

\section{Simplification of $H_{kk}^{(1)}$ under the linear order approximation}\label{appb}
In this appendix, we want to obtain the expression of Eq. (\ref{ohkk1}) under the first-order approximation of the quasistationary accreting process. Before analyzing the expression of $H_{kk}^{(i)}, i = 1, 2, 3$, four important identities that will be used heavily in the following calculations to simplify the expression of the equation should be introduced. Firstly, we should prove that two null vectors $k^a$ and $l^a$ on the background spacetime will satisfy the identity
\begin{equation}\label{knablaleq0}
	k^a \nabla_a l_b = 0\,.
\end{equation}
In order to demonstrate this identity, one can define the Killing vector $\xi^a = \left(\partial / \partial v \right)^a$ which generates the event horizon of the black holes on the background spacetime. Since the parameter $v$ is not the affine parameter, the geodesic equation for the Killing vector $\xi^a$ should be expressed as $\xi^a \nabla_a \xi^b = \kappa \xi^b$, where $\kappa$ represents the surface gravity of black holes. From the definition of $\xi^a$, the null vector $k^a$ can be obtained as $k^a = e^{- \kappa v} \xi^a$. Using the identity $k^a l_a = -1$, the second null vector $l^a$ can be naturally written as $l^a = e^{\kappa v} s^a$, where $s^a$ is also a Killing vector. Furthermore, two relationships between the two null Killing vectors can be given as
\begin{equation}
	\xi^a s_a = -1, \qquad \mathcal{L}_\xi s^a = \xi^b \nabla_b s^a - s^b \nabla_b \xi^a = 0\,.
\end{equation}
From the definitions of two Killing vectors, $k^a \nabla_a l^b$ on the background spacetime can be calculated as
\begin{equation}\label{knablal}
	\begin{split}
		k^a \nabla _a l^b = & e^{- \kappa v} \xi^a \nabla_a \left(e^{\kappa v} s^b \right) \\
		= & \xi^a \nabla_a s^b + \kappa s^b \\
		= &  s^a \nabla_a \xi^b + \kappa s^b \\
		= & \kappa s^a \hat{\epsilon}_{a}^{\ b} + \kappa s^b\\
		= & 0 \,,
	\end{split}
\end{equation}
where we have used the identity \cite{Iyer:1994ys}
\begin{equation}\label{defbinormal}
	\nabla_a \xi_b = \kappa \hat{\epsilon}_{ab}
\end{equation}
with the definition of the binormal $\hat{\epsilon}_{ab} = k_a \wedge l_b = \xi_a \wedge s_b$ in the last step. Furthermore, using the definition of $\xi^a$ and Eq. (\ref{defbinormal}), the second and third identities on the background spacetime can be respectively calculated as
\begin{equation}\label{gammanablabc}
	\begin{split}
		\gamma^{ab} \nabla_b k_c & = \gamma^{ab} \nabla_b \left(e^{- \kappa v} \xi_c \right) \\
		& = \gamma^{ab} \xi_c \nabla_b e^{- \kappa v} + \gamma^{ab} e^{- \kappa v} \nabla_b \xi_c \\
		& = \kappa \gamma^{ab} e^{- \kappa v} \hat{\epsilon}_{bc} \\
		& = 0\,,
	\end{split}
\end{equation}
and
\begin{equation}\label{gammanablacb}
	\begin{split}
		\gamma^{ab} \nabla_c k_b & = \gamma^{ab} \nabla_c \left(e^{- \kappa v} \xi_b \right)\\
		= & \gamma^{ab} \xi_b \nabla_c e^{- \kappa v} + \gamma^{ab} e^{- \kappa v} \nabla_c \xi_b \\
		= & \gamma^{ab} e^{- \kappa v} \kappa \hat{\epsilon}_{cb}\\
		= & 0\,.
	\end{split}
\end{equation}
From the results, one can see that two quantities $\gamma^{ab} \nabla_b k_c$ and $\gamma^{ab} \nabla_c k_b$ are both vanishing on the background spacetime. In other words, two quantities are both the first-order terms. The fourth identity on the background spacetime is
\begin{equation}\label{liekgammaab}
	\begin{split}
		\mathcal{L}_k \gamma_{ab} = & k^c \nabla_c \gamma_{ab} + \gamma_{cb} \nabla_a k^c + \gamma_{ac} \nabla_b k^c \\
		= & k^c \nabla_c \left(g_{ab} + k_a l_b + k_b l_a \right) \\
		= & l_b k^c \nabla_c k_a + k_a k^c \nabla_c l_b + l_a k^c \nabla_c k_b + k_b k^c \nabla_c l_a\\
		= & 0\,,
	\end{split}
\end{equation}
where we have used the definition of the induced metric and the results of Eq. (\ref{gammanablabc}) and Eq. (\ref{gammanablacb}) in the second step, the geodesic equation $k^b \nabla_b k_a = 0$ and the result of Eq. (\ref{knablal}) is used in the last step. Next, we will directly use the results of Eq. (\ref{knablaleq0}), Eq. (\ref{gammanablabc}), Eq. (\ref{gammanablacb}), and Eq. (\ref{liekgammaab}) to simplify the specific expression of $H_{kk}^{(i)}, i = 1, 2, 3$.

After expanding duplicated indexes of the second term in Eq. (\ref{ohkk1}), the specific expression of $H_{kk}^{(1)}$ can be written as
\begin{equation}\label{hkk1}
	\begin{split}
		H_{kk}^{(1)} = 2 R_{kk} F_{cd} F^{cd} - 4 F_{a}^{\ c} F_{b}^{\ d} k^a k^b R \gamma_{cd} - 2 k^a k^b \nabla_a \nabla_b \left(F^{cd} F_{cd} \right)\,.
	\end{split}
\end{equation}
Under the linear order approximation, the first term of Eq. (\ref{hkk1}) is
\begin{equation}
	2 R_{kk} \left(F_{cd} F^{cd} \right) = 2 \left(R_{kk} \right)_1 \left(F_{cd} F^{cd} \right)_0 \sim \mathcal{O} \left(\epsilon \right)\,.
\end{equation}
The second term of Eq. (\ref{hkk1}) is
\begin{equation}
	- 4 F_{a}^{\ c} F_{b}^{\ d} k^a k^b R \gamma_{cd} = - 4 \left(k^a F_{a}^{\ \hat{c}} \right)_1 \left(k^b F_{b \hat{c}} \right)_1 \left(R \right)_0 \simeq 0\,.
\end{equation}
The third of Eq. (\ref{hkk1}) can be simplified as
\begin{equation}
	- 2 k^a k^b \nabla_a \nabla_b \left(F^{cd} F_{cd} \right) = - 2 \left[\mathcal{L}_k^2 \left(F^{cd} F_{cd} \right) \right]_1 \sim \mathcal{O} \left(\epsilon \right)\,.
\end{equation}
Therefore, the expression of $H_{kk}^{(1)}$ under the first-order approximation can be simplified as
\begin{equation}
	\begin{split}
		H_{kk}^{(1)} = 2 R_{kk} F_{cd} F^{cd} - 2 \mathcal{L}_k^2 \left(F^{cd} F_{cd} \right)\,.
	\end{split}
\end{equation}

\section{Simplification of $H_{kk}^{(2)}$ under the linear order approximation}\label{appc}
We will continue to calculate Eq. (\ref{ohkk2}) in the third appendix. Expanding duplicated indexes in the expression again, Eq. (\ref{ohkk2}) can be further written as
\begin{equation}\label{hkk2}
	\begin{split}
		H_{kk}^{(2)} = & - 2 F_{ae} F_{bf} k^a k^b k^c k^d l^e l^f R_{cd} - 2 F_{a}^{\ c} F_{b}^{\ d} k^a k^b R^{ef} \gamma_{ce} \gamma_{df} \\
		& + 4 F_{ad} F_{b}^{\ e} k^a k^b k^c l^d R_{c}^{\ f} \gamma_{ef} + 4 k^a k^b k^c l^d \gamma^{ef} \nabla_b F_{ae} \nabla_c F_{df}  \\
		& + 4 F_{ad} F_{b}^{\ e} k^a k^b k^c l^d R_{c}^{\ f} \gamma_{ef} + 2 F_{a}^{\ e} k^a k^b k^c l^d \gamma_{e}^{\ f} \nabla_c \nabla_b F_{df} \\
		& + 4 k^a k^b k^c k^d l^e l^f \nabla_b F_{ae} \nabla_d F_{cf} + 4 F_{ae} k^a k^b k^c k^d l^e l^f \nabla_d \nabla_c F_{bf} \\
		& + 4 F_{a}^{\ e} k^a k^b k^c l^d \gamma_{e}^{f} \nabla_d \nabla_c F_{bf} + 2 k^a k^b \gamma^{cd} \gamma^{ef} \nabla_b F_{df} \nabla_e F_{ac} \\
		& + 4 F_{a}^{\ e} k^a k^b k^c l^d \gamma_{e}^{f} \nabla_d \nabla_c F_{bf} + 2 F_{a}^{\ c} k^a k^b \gamma_{c}^{\ d} \nabla_e \nabla^e F_{bd} \\
		& + 2 k^a k^b \gamma^{cd} \gamma^{ef} \nabla_e F_{ac} \nabla_f F_{bd} - 2 k^a k^b k^c l^d \gamma^{ef} \nabla_b F_{ad} \nabla_f F_{ce} \\
		& + 2 k^a k^b \gamma^{cd} \gamma^{ef} \nabla_b F_{ac} \nabla_f F_{de} + 2 F^{cd} k^a k^b \gamma_{c}^{\ e} \gamma_{d}^{\ f} \nabla_f \nabla_b F_{ae} \\
		& + 2 F_{a}^{\ c} k^a k^b \gamma_{c}^{\ d} \gamma^{ef} \nabla_f \nabla_b F_{de} - 2 F_{a}^{\ e} k^a k^b k^c l^d \gamma_{e}^{\ f} \nabla_f \nabla_c F_{bd} \\
		& - 2 F_{ad} k^a k^b k^c l^d \gamma^{ef} \nabla_f \nabla_c F_{be}\,.
	\end{split}
\end{equation}
For the first term of Eq. (\ref{hkk2}), it can be given as
\begin{equation}
	\begin{split}
		- 2 F_{ae} F_{bf} k^a k^b k^c k^d l^e l^f R_{cd} = - 2 \left(k^a F_{ae} l^e \right)_0 \left(k^b F_{bf} l^f \right)_0 \left(R_{kk} \right)_1 \sim \mathcal{O} \left(\epsilon \right)\,.
	\end{split}
\end{equation}
The second term of Eq. (\ref{hkk2}) is
\begin{equation}
	\begin{split}
		- 2 F_{a}^{\ c} F_{b}^{\ d} k^a k^b R^{ef} \gamma_{ce} \gamma_{df} = - 2 \left(k^a F_{a \hat{c}} \right)_1 \left(k^b F_{b \hat{d}} \right)_1 \left(R^{\hat{c} \hat{d}} \right)_0 \simeq 0\,.
	\end{split}
\end{equation}
The third term of Eq. (\ref{hkk2}) is given as
\begin{equation}
	\begin{split}
		4 F_{ad} F_{b}^{\ e} k^a k^b k^c l^d R_{c}^{\ f} \gamma_{ef} = 4 \left(k^a F_{ad} l^d \right)_0 \left(k^b F_{b \hat{e}} \right)_1 \left(k^c R_{c}^{\ \hat{e}} \right)_1 \simeq 0\,.
	\end{split}
\end{equation}
The fourth term of Eq. (\ref{hkk2}) is
\begin{equation}
	\begin{split}
		& 4 k^a k^b k^c l^d \gamma^{ef} \nabla_b F_{ae} \nabla_c F_{df} \\
		= & 4 k^b \nabla_b \left(k^a F_{ae} \gamma^{e}_{\ g} \right) k^c \nabla_c \left(l^d F_{df} \gamma^{fg} \right) \\
		= & 4 \left[\left[\mathcal{L}_k \left(k^a F_{ae} \gamma^{e}_{\ g} \right) \right]_1 - \left(k^a F_{ae} \gamma^{e}_{\ m} \right)_1 \left(\gamma^{m}_{\ b} \nabla_g k^b \right)_1 \right] \\
		& \times \left[\left[\mathcal{L}_k \left(l^d F_{df} \gamma^{fg} \right) \right]_0 + \left(l^d F_{df} \gamma^{f}_{\ n} \right)_0 \left(\gamma^{nc} \nabla_c k^g \right)_1 \right] \\
		\simeq & 4 \left[\mathcal{L}_k \left(k^a F_{a \hat{e}} \right) \right]_1 \left[\mathcal{L}_k \left(l^d F_{d}^{\ \hat{e}} \right) \right]_0 \\
		\sim & \mathcal{O} \left(\epsilon \right)\,.
	\end{split}
\end{equation}
The fifth term of Eq. (\ref{hkk2}) can be written as
\begin{equation}
	\begin{split}
		& 2 F_{d}^{\ e} k^a k^b k^c l^d \gamma_{e}^{\ f} \nabla_c \nabla_b F_{af} \\
		= & 2 \left(l^d F_{d}^{\ \hat{g}} \right)_0 k^c \nabla_c \left[k^b \nabla_b \left(k^a F_{af} \gamma^{f}_{\ g} \right) \right] \\
		= & 2 \left(l^d F_{d}^{\ \hat{g}} \right)_0 k^c \nabla_c \left[\mathcal{L}_k \left(k^a F_{a\hat{g}} \right) - \left(k^a F_{af} \gamma^{f}_{\ h} \right)_1 \left(\gamma^{h}_{\ b} \nabla_{\hat{g}} k^b \right)_1 \right] \\
		\simeq & 2 \left(l^d F_{d}^{\ \hat{g}} \right)_0 \left[\mathcal{L}_k^2 \left(k^a F_{a\hat{g}} \right) - \left[\mathcal{L}_k \left(k^a F_{a\hat{b}} \right) \right]_1 \left(\gamma^{\hat{b}}_{\ c} \nabla_{\hat{g}} k^c \right)_1 \right] \\
		\simeq & 2 \left(l^d F_{d}^{\ \hat{e}} \right)_0 \left[\mathcal{L}_k^2 \left(k^a F_{a \hat{e}} \right) \right]_1\\
		\sim & \mathcal{O} \left(\epsilon \right)\,.
	\end{split}
\end{equation}
The sixth term of Eq. (\ref{hkk2}) is given as
\begin{equation}
	\begin{split}
		& 2 F_{a}^{\ e} k^a k^b k^c l^d \gamma_{e}^{\ f} \nabla_c \nabla_b F_{df} \\
		= & 2 \left(k^a F_{a}^{\ \hat{g}} \right)_1 k^c \nabla_c \left[k^b \nabla_b \left(l^d F_{df} \gamma^{f}_{\ g} \right) \right] \\
		= & 2 \left(k^a F_{a}^{\ \hat{g}} \right)_1 k^c \nabla_c \left[\mathcal{L}_k \left(l^d F_{d\hat{g}} \right) - \left(l^d F_{d\hat{h}} \right)_0 \left(\gamma^{\hat{h}}_{\ b} \nabla_g k^b \right)_1 \right] \\
		\simeq & 2 \left(k^a F_{a}^{\ \hat{g}} \right)_1 \left[\mathcal{L}_k^2 \left(l^d F_{d\hat{g}} \right) - \left[\mathcal{L}_k \left(l^d F_{d\hat{b}} \right) \right]_0 \left(\gamma^{\hat{b}}_{\ c} \nabla_g k^c \right)_1 \right] \\
		\simeq & 2 \left(k^a F_{a \hat{e}} \right)_1 \left[\mathcal{L}_k^2 \left(l^d F_{d}^{\ \hat{e}} \right) \right]_0 \\
		\sim & \mathcal{O} \left(\epsilon \right)\,.
	\end{split}
\end{equation}
The seventh term of Eq. (\ref{hkk2}) can be expressed as
\begin{equation}
	\begin{split}
		4 k^a k^b k^c k^d l^e l^f \nabla_b F_{ae} \nabla_d F_{cf} = 4 \left[\mathcal{L}_k \left(k^a F_{ae} l^e \right) \right]_1 \left[\mathcal{L}_k \left(k^c F_{cf} l^f \right) \right]_1 \simeq 0\,.
	\end{split}
\end{equation}
The eighth term of Eq. (\ref{hkk2}) is written as
\begin{equation}
	\begin{split}
		4 F_{ae} k^a k^b k^c k^d l^e l^f \nabla_d \nabla_c F_{bf} = 4 \left(k^a F_{ae} l^e \right)_0 \left[\mathcal{L}_k^2 \left(k^b F_{bf} l^f \right) \right]_1 \sim \mathcal{O} \left(\epsilon \right)\,.
	\end{split}
\end{equation}
Utilizing the following relation
\begin{equation}
	\left(\nabla_d \nabla_c - \nabla_c \nabla_d \right) F_{bf} = R_{dcb}^{\ \ \ g} F_{gf} + R_{dcf}^{\ \ \ g} F_{bg}\,,
\end{equation}
one can write the ninth term of Eq. (\ref{hkk2}) as
\begin{equation}
	\begin{split}
		& 4 F_{a}^{\ e} k^a k^b k^c l^d \gamma_{e}^{f} \nabla_d \nabla_c F_{bf} \\
		= & 4 F_{ae} k^a k^b k^c l^d \gamma^{ef} \nabla_c \nabla_d F_{bf} + 4 F_{ae} k^a k^b k^c l^d \gamma^{ef} R_{dcb}^{\ \ \ g} F_{gf} \\
		& + 4 F_{ae} k^a k^b k^c l^d \gamma^{ef} R_{dcf}^{\ \ \ g} F_{bg}\,.
	\end{split}
\end{equation}
According to the Bianchi identity of the Maxwell field $\nabla_{[d} F_{bf]} = 0$, the ninth term can further be written as
\begin{equation}\label{simpninthterm}
	\begin{split}
		& 4 F_{a}^{\ e} k^a k^b k^c l^d \gamma_{e}^{f} \nabla_d \nabla_c F_{bf} \\
		= & - 4 F_{ae} k^a k^b k^c l^d \gamma^{ef} \nabla_c \nabla_f F_{db} - 4 F_{ae} k^a k^b k^c l^d \gamma^{ef} \nabla_c \nabla_b F_{fd} \\
		& + 4 F_{ae} k^a k^b k^c l^d \gamma^{ef} R_{dcb}^{\ \ \ g} F_{gf} + 4 F_{ae} k^a k^b k^c l^d \gamma^{ef} R_{dcf}^{\ \ \ g} F_{bg}\,.
	\end{split}
\end{equation}
The first term of Eq. (\ref{simpninthterm}) is given as
\begin{equation}\label{firstsimpninthterm}
	\begin{split}
		& - 4 F_{ae} k^a k^b k^c l^d \gamma^{ef} \nabla_c \nabla_f F_{db} \\
		= & - 4 k^a F_{ae} \gamma^{ef} k^c \nabla_c \left(k^b l^d \nabla_f F_{db} \right) \\
		= & - 4 k^a F_{ae} \gamma^{ef} k^c \nabla_c \left[\nabla_f \left(k^b F_{db} l^d \right) - \left(\nabla_f k^b \right) l^d F_{db} - k^b \left(\nabla_f l^d \right) F_{db} \right] \\
		= & - 4 k^a F_{ae} \gamma^{ef} k^c \nabla_c \nabla_f \left(k^b F_{db} l^d \right) + 4 k^a F_{ae} \gamma^{ef} k^c \nabla_c \left(l^d F_{db} \right) \nabla_f k^b \\
		& + 4 k^a F_{ae} \gamma^{ef} l^d F_{db} k^c \nabla_c \left(\nabla_f k^b \right) + 4 k^a F_{ae} \gamma^{ef} k^c \nabla_c \left(F_{db} k^b \right) \nabla_f l^d \\
		& + 4 k^a F_{ae} \gamma^{ef} F_{db} k^b k^c \nabla_c \left(\nabla_f l^d \right)\,.
	\end{split}
\end{equation}
For the first term of Eq. (\ref{firstsimpninthterm}), it can be calculated as
\begin{equation}
	\begin{split}
		& - 4 k^a F_{ae} \gamma^{ef} k^c \nabla_c \nabla_f \left(k^b F_{db} l^d \right) \\
		= & - 4 \left(k^a F_{ae} \gamma^{eg} \right) \gamma_{g}^{\ f} k^c \nabla_f \nabla_c \left(k^b F_{db} l^d \right) \\
		= & - 4 \left(k^a F_{a \hat{e}} \right)_1 D^{\hat{e}} \left[\mathcal{L}_k \left(k^b F_{db} l^d \right) \right]_1 - 4 \left(k^a F_{a}^{\ \hat{e}} \right)_1 \left(\gamma_{\hat{e}}^{\ f} \nabla_f k^c \right)_1 \nabla_c \left(k^b F_{db} l^d \right) \\
		\simeq & 0\,.
	\end{split}
\end{equation}
The second of Eq. (\ref{firstsimpninthterm}) is
\begin{equation}
	\begin{split}
		4 k^a F_{ae} \gamma^{ef} k^c \nabla_c \left(l^d F_{db} \right) \nabla_f k^b = 4 \left(k^a F_{a}^{\ \hat{e}} \right)_1 \left(\gamma_{\hat{e}}^{\ f} \nabla_f k^b \right)_1 k^c \nabla_c \left(l^d F_{db} \right) \simeq 0\,.
	\end{split}
\end{equation}
The third term of Eq. (\ref{firstsimpninthterm}) can be evaluated as
\begin{equation}
	\begin{split}
		4 k^a F_{ae} \gamma^{ef} l^d F_{db} k^c \nabla_c \left(\nabla_f k^b \right)
		\simeq 4 \left(k^a F_{a}^{\ \hat{e}} \right)_1 \left[\mathcal{L}_k \left(\gamma_{\hat{e}}^{\ f} \nabla_f k^b \right) \right]_1 \left(l^d F_{db} \right) \simeq 0\,.
	\end{split}
\end{equation}
For the fourth term of Eq. (\ref{firstsimpninthterm}), it can be calculated as
\begin{equation}
	\begin{split}
		& 4 k^a F_{ae} \gamma^{ef} k^c \nabla_c \left(F_{db} k^b \right) \nabla_f l^d \\
		= & 4 \left(k^a F_{a}^{\ \hat{h}} \right) k^c \nabla_c \left(\gamma_{d}^{\ g} F_{gb} k^b - k_d l^g F_{gb} k^b - k^{(g} k^{d)} F_{[gb]} l_d \right) \left(\nabla_{\hat{h}} l^d \right) \\
		= & 4 \left(k^a F_{a}^{\ \hat{h}} \right)_1 \left[\mathcal{L}_k \left(F_{\hat{d} b} k^b \right) \right]_1 \left(\nabla_{\hat{h}} l^{\hat{d}} \right) - 4 \left(k^a F_{a}^{\ \hat{h}} \right)_1 \left(F_{\hat{m} b} k^b \right)_1 \left(\gamma^{\hat{m}}_{\ c} \nabla_d k^c \right)_1 \left(\nabla_{\hat{h}} l^{d} \right)\\
		& - 4 \left(k^a F_{a}^{\ \hat{h}} \right)_1 \left[\mathcal{L}_k \left(l^g F_{gb} k^b \right) \right]_1 \left(k_d \nabla_{\hat{h}} l^d \right) \\
		\simeq & 0\,.
	\end{split}
\end{equation}
For the fifth term of Eq. (\ref{firstsimpninthterm}), we have
\begin{equation}
	\begin{split}
		& 4 k^a F_{ae} \gamma^{ef} F_{db} k^b k^c \nabla_c \left(\nabla_f l^d \right) \\
		= & 4 \left(k^a F_{a \hat{e}} \right)_1 \left(\gamma^{g}_{\ d} F_{gb} k^b \right)_1 k^c \nabla_c \left(\nabla^{\hat{e}} l^d \right) - 4 \left(k^a F_{a \hat{e}} \right)_1 \left(k^{(g} k^{b)} F_{[gb]} \right) k^c \nabla_c \left(l_d \nabla^{\hat{e}} l^d \right) \\
		& + 4 \left(k^a F_{a}^{\ \hat{e}} \right)_1 \left(l^g F_{gb} k^b \right)_0 k^c \nabla_c \left[l^d \left(\gamma_{\hat{e}}^{\ f} \nabla_f k_d \right)_1 \right]\\
		\simeq & 0\,.
	\end{split}
\end{equation}
The result of the first term of Eq. (\ref{simpninthterm}) under the first-order approximation is
\begin{equation}
	- 4 k^a F_{ae} \gamma^{ef} k^c \nabla_c \nabla_f \left(k^b F_{db} l^d \right) \simeq 0\,.
\end{equation}
The second term of Eq. (\ref{simpninthterm}) is
\begin{equation}
	\begin{split}
		& - 4 F_{ae} k^a k^b k^c l^d \gamma^{ef} \nabla_c \nabla_b F_{fd} \\
		= & - 4 \left(k^a F_{ae} \gamma^{eg} \right) k^c \nabla_c \left[k^b \nabla_b \left(\gamma_{g}^{\ f} F_{fd} l^d \right) \right] \\
		\simeq & - 4 \left(k^a F_{a}^{\ \hat{e}} \right)_1 \left[ \mathcal{L}_k^2 \left(F_{\hat{e}d} l^d \right) \right]_0 \\
		\sim & \mathcal{O} \left(\epsilon \right)\,.
	\end{split}
\end{equation}
The third term of Eq. (\ref{simpninthterm}) is calculated as
\begin{equation}
	\begin{split}
		& 4 F_{ae} k^a k^b k^c l^d \gamma^{ef} R_{dcb}^{\ \ \ g} F_{gf} \\
		= & 4 \left(k^a F_{a \hat{e}} \right)_1 \left(k^b k^c l^d R_{acb \hat{g}} \right)_1 \left(F^{\hat{g} \hat{e}} \right)_0 - 4 \left(k^a F_{a \hat{e}} \right) \left(k^c l^d k^{(b} k^{h)} R_{dc[bh]} \right) \left(l^g F_{g}^{\ \hat{e}} \right)\\
		& - 4 \left(k^a F_{a \hat{e}} \right)_1 \left(k^b k^c l^d l^h R_{dcbh} \right)_0 \left(k^g F_{g}^{\ \hat{e}} \right)_1 \\
		\simeq & 0\,.
	\end{split}
\end{equation}
The fourth term of Eq. (\ref{simpninthterm}) is written as
\begin{equation}
	\begin{split}
		& 4 F_{ae} k^a k^b k^c l^d \gamma^{ef} R_{dcf}^{\ \ \ g} F_{bg} \\
		= & 4 \left(k^a F_{a}^{\ \hat{e}} \right)_1 \left(k^c l^d R_{dc \hat{e} \hat{g}} \right)_0 \left(k^b F_{b}^{\ \hat{g}} \right)_1 - 4 \left(k^a F_{a}^{\ \hat{e}} \right)_1 \left(k^{c} k^h l^{d} R_{dc \hat{e} h}\right)_1 \left(k^b F_{bg} l^g \right)_0 \\
		& - 4 \left(k^a F_{a}^{\ \hat{e}} \right) \left(k^c l^d l^h R_{dc\hat{e}h}\right) \left(k^{(b} k^{g)} F_{[bg]} \right) \\
		\simeq & 0\,.
	\end{split}
\end{equation}
So the ninth term of Eq. (\ref{hkk2}) under the first-order approximation can be expressed as
\begin{equation}
	4 F_{a}^{\ e} k^a k^b k^c l^d \gamma_{e}^{f} \nabla_d \nabla_c F_{bf} \simeq - 4 \left(k^a F_{a}^{\ \hat{e}} \right) \left[ \mathcal{L}_k^2 \left(F_{\hat{e}d} l^d \right) \right]\,.
\end{equation}
The tenth term of Eq. (\ref{hkk2}) is
\begin{equation}
	\begin{split}
		& 2 k^a k^b \gamma^{cd} \gamma^{ef} \nabla_b F_{df} \nabla_e F_{ac} \\
		= & 2 \left(k^b \nabla_b F_{df} \right) \gamma^{cd} \gamma^{ef} \nabla_e \left(k^a F_{ac} \right) - 2 \left(k^b \nabla_b F_{df} \right) \gamma^{cd} F_{ac} \left(\gamma^{ef} \nabla_e k^a  \right) \\
		= & 2 \left(k^b \nabla_b F_{df} \right) \gamma^{cd} \gamma^{ef} \nabla_e \left(k^a F_{ag} \gamma^{g}_{\ c} - k^{(a} k^{g)} F_{[ag]} l_c - k_c k^a F_{ag} l^g \right) \\
		& - 2 \left[\mathcal{L}_k \left(F_{\hat{m} \hat{n}} \right) \right]_1 \left(F_{a}^{\ \hat{m}} \right) \left(\gamma^{e \hat{n}} \nabla_e k^a \right)_1 + 2 \left(F_{d \hat{n}} \right) \left(\gamma_{b}^{\ d}\nabla_{\hat{m}} k^b \right)_1 \left(F_{a}^{\ \hat{m}} \right) \left(\gamma^{e \hat{n}} \nabla_e k^a \right)_1\\
		& + 2 \left(F_{\hat{m}f} \right) \left(\gamma_{b}^{\ f} \nabla_n k^b \right)_1 \left(F_{a}^{\ \hat{m}} \right) \left(\gamma^{en} \nabla_e k^a \right)_1\\
		\simeq & 2 \left[\mathcal{L}_k \left(F^{\hat{c} \hat{e}} \right) \right]_1 D_{\hat{e}} \left(k^a F_{a \hat{c}} \right)_1 - 2 \left[\mathcal{L}_k \left(F_{\hat{m} \hat{n}} \right) \right]_1 \left(k^a F_{ag} l^g\right)_0 \left(\gamma^{c \hat{m}} \gamma^{e \hat{n}} \nabla_e k_c \right)_1 \\
		& - 2 \left(k^b \nabla_b F_{df} \right) \left(\gamma^{cd} k_c \right) D^f \left(k^a F_{ag} l^g\right) \\
		\simeq & 0\,.
	\end{split}
\end{equation}
The eleventh term of Eq. (\ref{hkk2}) is further calculated as
\begin{equation}
	\begin{split}
		& - 2 k^a k^b k^c l^d \gamma^{ef} \nabla_c F_{bf} \nabla_e F_{ad} \\
		= & - 2 \left[k^c \nabla_c \left(k^b F_{bf} \gamma^{f}_{\ g}\right) \right] \left(k^a l^d \gamma^{ge} \nabla_e F_{ad} \right) \\
		= & - 2 \left[\left[\mathcal{L}_k \left(k^b F_{b \hat{g}} \right) \right]_1 - \left(k^b F_{b \hat{f}} \right)_1 \left(\gamma^{\hat{f}}_{\ c} \nabla_{\hat{g}} k^c \right)_1 \right] \\
		& \times \left[D^{\hat{g}} \left(k^a F_{ad} l^d \right) - \left(F_{ad} l^d \right) \left(\gamma^{\hat{g}e} \nabla_e k^a \right)_1 - \left(k^a F_{ad} \right) \left(\gamma^{\hat{g}e} \nabla_e l^d \right) \right] \\
		\simeq & - 2 \left[\mathcal{L}_k \left(k^b F_{b \hat{e}} \right) \right]_1 \left[D^{\hat{e}} \left(k^a F_{ad} l^d \right) \right]_0 + 2 \left[\mathcal{L}_k \left(k^b F_{b\hat{e}} \right) \right]_1 \left(k^a F_{a \hat{d}} \right)_1 \left(\nabla^{\hat{e}} l^{\hat{d}} \right)\\
		& - 2 \left[\mathcal{L}_k \left(k^b F_{b \hat{e}} \right) \right] \left(k^{(a} k^{c)} F_{[ac]} \right) \left(l_d \nabla^{\hat{e}} l^d \right) + 2 \left[\mathcal{L}_k \left(k^b F_{b \hat{e}} \right) \right]_1 \left(k^a F_{ac} l^c \right)_0 \left(l^d \nabla^{\hat{e}} k_d \right)_1 \\
		\simeq & - 2 \left[\mathcal{L}_k \left(k^b F_{b}^{\ \hat{e}} \right) \right]_1 \left[D_{\hat{e}} \left(k^a F_{ad} l^d \right) \right]_0\\
		\sim & \mathcal{O} \left(\epsilon \right)\,.
	\end{split}
\end{equation}
The twelfth term of Eq. (\ref{hkk2}) is given as
\begin{equation}\label{twelfthhkk2}
	\begin{split}
		& 2 F_{a}^{\ c} k^a k^b \gamma_{c}^{\ d} \nabla_e \nabla^e F_{bd} \\
		= & - 2 F_{ae} k^a k^b k^c l^d \gamma^{ef} \nabla_c \nabla_d F_{bf} - 2 F_{ae} k^a k^b k^c l^d \gamma^{ef} \nabla_d \nabla_c F_{bf} \\
		& + 2 F_{ac} k^a k^b \gamma^{cd} \gamma^{ef} \nabla_f \nabla_e F_{bd}\,.
	\end{split}
\end{equation}
Using the Bianchi identity again, the first term of Eq. (\ref{twelfthhkk2}) can be expressed as
\begin{equation}\label{firsttwelfthhkk2}
	\begin{split}
		& - 2 F_{ae} k^a k^b k^c l^d \gamma^{ef} \nabla_c \nabla_d F_{bf} \\
		= & 2 k^a F_{ae} \gamma^{ef} k^c k^b l^d \nabla_c \nabla_f F_{db} + 2 k^a F_{ae} \gamma^{ef} k^c k^b l^d \nabla_c \nabla_b F_{fd}\,.
	\end{split}
\end{equation}
For the first term of Eq. (\ref{firsttwelfthhkk2}), it can be calculated as
\begin{equation}
	\begin{split}
		& 2 k^a F_{ae} \gamma^{ef} k^c k^b l^d \nabla_c \nabla_f F_{db} \\
		= & 2 \left(k^a F_{ae} \gamma^{ef} \right) k^c \nabla_c \left[\nabla_f \left(k^b F_{db} l^d \right) - l^d F_{db} \nabla_f k^b - k^b F_{db} \nabla_f l^d \right] \\
		= & 2 \left(k^a F_{a}^{\ \hat{g}} \right)_1 k^c \nabla_c \left[\gamma_{\hat{g}}^{\ f} \nabla_f \left(k^b F_{db} l^d \right) - l^d F_{db} \left(\gamma_{\hat{g}}^{\ f} \nabla_f k^b \right)_1 - k^b F_{db} \gamma_{\hat{g}}^{\ f} \nabla_f l^d \right] \\
		\simeq & 2 \left(k^a F_{a}^{\ \hat{g}} \right)_1 k^c \nabla_c \left[\gamma_{\hat{g}}^{\ f} \nabla_f \left(k^b F_{db} l^d \right) \right] - 2 \left(k^a F_{a}^{\ \hat{g}} \right)_1 k^c \nabla_c \left[\left(k^b F_{\hat{d} b} \right)_1 \left(\nabla_{\hat{g}} l^{\hat{d}} \right) \right]\\
		& + 2 \left(k^a F_{a}^{\ \hat{g}} \right)_1 k^c \nabla_c \left(k^{(b} k^{h)} F_{[hb]} l_d \gamma_{\hat{g}}^{\ f} \nabla_f l^d \right) - 2 \left(k^a F_{a}^{\ \hat{g}} \right)_1 k^c \nabla_c \left[\left(k^b F_{hb} l^h \right) \left(l^d \nabla_{\hat{g}} k_d \right)_1 \right]\\
		\simeq & 2 \left(k^a F_{a}^{\ \hat{g}} \right)_1 \left[\mathcal{L}_k \left[D_{\hat{g}} \left(k^b F_{db} l^d \right) \right] - \left(\gamma_{c}^{\ f} \nabla_{\hat{g}} k^c \right)_1 \nabla_f \left(k^b F_{db} l^d \right) \right]\\
		\simeq & 2 \left(k^a F_{a}^{\ \hat{e}} \right)_1 D_{\hat{e}} \left[\mathcal{L}_k \left(k^b F_{db} l^d \right) \right]_1 \\
		\simeq & 0\,.
	\end{split}
\end{equation}
The second term of Eq. (\ref{firsttwelfthhkk2}) is
\begin{equation}
	\begin{split}
		& 2 k^a F_{ae} \gamma^{ef} k^c k^b l^d \nabla_c \nabla_b F_{fd} \\
		= & 2 \left(k^a F_{ae} \gamma^{eg} \right)_1 k^c \nabla_c \left[k^b \nabla_b \left(\gamma_{g}^{\ f} F_{fd} l^d \right) \right] \\
		\simeq & 2 \left(k^a F_{a}^{\ \hat{e}} \right)_1 \left[\mathcal{L}_k^2 \left(F_{\hat{e}d} l^d \right) \right]_0\\
		\sim & \mathcal{O} \left(\epsilon \right)\,.
	\end{split}
\end{equation}
The result of the first term of Eq. (\ref{twelfthhkk2}) is
\begin{equation}
	- 2 F_{ae} k^a k^b k^c l^d \gamma^{ef} \nabla_c \nabla_d F_{bf} \simeq 2 \left(k^a F_{a}^{\ \hat{e}} \right) \left[\mathcal{L}_k^2 \left(F_{\hat{e}d} l^d \right) \right]\,.
\end{equation}
According to the definition of the Riemann tensor
\begin{equation}\label{riemanntensor}
	\left(\nabla_a \nabla_b - \nabla_b \nabla_a \right) X_{cd} = R_{abc}^{\ \ \ e} X_{ed} + R_{abd}^{\ \ \ e} X_{ce}\,,
\end{equation}
the second term of Eq. (\ref{twelfthhkk2}) is calculated as
\begin{equation}
	\begin{split}\label{secondtwelfthhkk2}
		& - 2 F_{ae} k^a k^b k^c l^d \gamma^{ef} \nabla_d \nabla_c F_{bf} \\
		= & - 2 F_{ae} k^a k^b k^c l^d \gamma^{ef} \left(\nabla_c \nabla_d F_{bf} + R_{dcb}^{\ \ \ g} F_{gf} + R_{dcf}^{\ \ \ g} F_{bg} \right) \\
		= & - 2 F_{ae} k^a k^b k^c l^d \gamma^{ef} \nabla_c \nabla_d F_{bf} - 2 F_{ae} k^a k^b k^c l^d \gamma^{ef} R_{dcb}^{\ \ \ g} F_{gf} \\
		& - 2 F_{ae} k^a k^b k^c l^d \gamma^{ef} R_{dcf}^{\ \ \ g} F_{bg}\,.
	\end{split}
\end{equation}
Since the first term of Eq. (\ref{secondtwelfthhkk2}) is identical with the first term of Eq. (\ref{firsttwelfthhkk2}), the result of this term can be directly written as
\begin{equation}
	\begin{split}
		- 2 F_{ae} k^a k^b k^c l^d \gamma^{ef} \nabla_c \nabla_d F_{bf} \simeq 2 \left(k^a F_{a}^{\ \hat{e}} \right) \left[\mathcal{L}_k^2 \left(F_{\hat{e}d} l^d \right) \right]\,.
	\end{split}
\end{equation}
The second and the third term of Eq. (\ref{secondtwelfthhkk2}) can be further calculated as
\begin{equation}
	\begin{split}
		& - 2 F_{ae} k^a k^b k^c l^d \gamma^{ef} R_{dcb}^{\ \ \ g} F_{gf} - 2 F_{ae} k^a k^b k^c l^d \gamma^{ef} R_{dcf}^{\ \ \ g} F_{bg} \\
		= & - 2 \left(k^a F_{a \hat{e}} \right) _1 \left(F^{\hat{f} \hat{e}} \right)_0 \left(k^b k^c l^d R_{dcb\hat{f}} \right)_1 - 2 \left(k^a F_{a\hat{e}} \right)_1 \left(k^g F_{g}^{\ \hat{e}} \right)_1 \left(k^b k^c l^d l^h R_{dcbh} \right)_0 \\
		& - 2 \left(k^a F_{a}^{\ \hat{e}} \right)_1 \left(k^b F_{b}^{\ \hat{f}} \right)_1 \left(k^c l^d R_{dc\hat{e}\hat{f}} \right)_0 - 2 \left(k^a F_{a}^{\ \hat{e}} \right)_1 \left(k^b l^g F_{bg} \right)_0 \left(k^c l^d k^h R_{dc\hat{e}h} \right)_1\\
		& - 2 \left(k^a F_{a\hat{e}} \right) \left(l^g F_{g}^{\ \hat{e}} \right) \left(k^c l^d k^{(b} k^{h)} R_{dc[bh]} \right) - 2 \left(k^a F_{a}^{\ \hat{e}} \right) \left(k^{(b} k^{g)} F_{[bg]} \right) \left(k^c l^d l^h R_{dc\hat{e}h} \right)\\
		\simeq & 0\,.
	\end{split}
\end{equation}
Therefore, the second term of Eq. (\ref{secondtwelfthhkk2}) under the linear order approximation is
\begin{equation}
	- 2 F_{ae} k^a k^b k^c l^d \gamma^{ef} \nabla_d \nabla_c F_{bf} \simeq 2 \left(k^a F_{a}^{\ \hat{e}} \right) \left[\mathcal{L}_k^2 \left(F_{\hat{e}d} l^d \right) \right] \sim \mathcal{O} \left(\epsilon \right)\,.
\end{equation}
The third term of Eq. (\ref{twelfthhkk2}) is further written as
\begin{equation}
	\begin{split}
		& 2 F_{ac} k^a k^b \gamma^{cd} \gamma^{ef} \nabla_f \nabla_e F_{bd} \\
		= & 2 \left(k^a F_{a}^{\ \hat{g}} \right)_1 \gamma^{hf} \nabla_f \left(\gamma_{\hat{g}}^{\ d} \gamma_{h}^{\ e} k^b \nabla_e F_{bd} \right)
		- 2 \left(k^a F_{a}^{\ \hat{g}} \right)_1 \left(\gamma_{\hat{g}}^{\ d} \gamma_{h}^{\ e} \nabla_e F_{bd} \right) \left(\gamma^{hf} \nabla_f k^b \right)_1 \\
		\simeq & 2 \left(k^a F_{a}^{\ \hat{g}} \right)_1 D_f \left[D^f \left(k^b F_{b\hat{g}} \right)_1 - \gamma^{d}_{\ \hat{g}} F_{bd} \left(\gamma^{fe} \nabla_e k^b \right)_1 \right] \\
		\simeq & 0\,.
	\end{split}
\end{equation}
The result of twelfth term of Eq. (\ref{hkk2}) under the first-order approximation is given as
\begin{equation}
	\begin{split}
		2 F_{a}^{\ c} k^a k^b \gamma_{c}^{\ d} \nabla_e \nabla^e F_{bd} \simeq 4 \left(k^a F_{a}^{\ \hat{e}} \right) \left[\mathcal{L}_k^2 \left(F_{\hat{e}d} l^d \right) \right]\,.
	\end{split}
\end{equation}
The thirteenth term of Eq. (\ref{hkk2}) can be calculated as
\begin{equation}
	\begin{split}
		& 2 k^a k^b \gamma^{cd} \gamma^{ef} \nabla_e F_{ac} \nabla_f F_{bd} \\
		= & 2 \left[D^h \left(k^a F_{a}^{\ \hat{g}} \right)_1 - F_{a}^{\ \hat{g}} \left(\gamma^{eh} \nabla_e k^a \right)_1 \right] \left[D_h \left(k^b F_{b \hat{g}} \right)_1 - F_{b \hat{g}} \left(\gamma_{h}^{f} \nabla_f k^b \right)_1 \right] \\
		\simeq & 0\,.
	\end{split}
\end{equation}
The fourteenth term of Eq. (\ref{hkk2}) can be expressed as
\begin{equation}
	\begin{split}
		& - 2 k^a k^b k^c l^d \gamma^{ef} \nabla_b F_{ad} \nabla_f F_{ce} \\
		= & - 2 \left[\mathcal{L}_k \left(k^a F_{ad} l^d \right) \right]_1 \left[D^{\hat{g}} \left(k^c F_{c \hat{g}} \right)_1 - F_{c \hat{g}} \left(\gamma^{\hat{g}f} \nabla_f k^c \right)_1 \right] \\
		\simeq & 0\,.
	\end{split}
\end{equation}
The fifteenth term of Eq. (\ref{hkk2}) is given as
\begin{equation}
	\begin{split}
		& 2 k^a k^b \gamma^{cd} \gamma^{ef} \nabla_b F_{ac} \nabla_f F_{de} \\
		= & 2 k^b \nabla_b \left(k^a F_{ac} \gamma^{c}_{\ g} \right) \gamma^{gd} \gamma^{e}_{\ h} \gamma^{hf} \nabla_f F_{de} \\
		= & 2 \left[\mathcal{L}_k \left(k^a F_{a\hat{g}} \right) - \left(k^a F_{a\hat{m}} \right)_1 \left(\gamma^{\hat{m}}_{\ b} \nabla_{\hat{g}} k^b \right)_1 \right] \left[\gamma^{hf} \nabla_f \left(F^{\hat{g}}_{\ h} \right) \right] \\
		\simeq & 2 \left[\mathcal{L}_k \left(k^a F_{a\hat{c}} \right) \right]_1 \left[D_e \left( F^{\hat{c} \hat{e}} \right) \right]_0\\
		\sim & \mathcal{O} \left(\epsilon \right)\,.
 	\end{split}
\end{equation}
The sixteenth term of Eq. (\ref{hkk2}) is
\begin{equation}\label{hkk2sixteenth}
	\begin{split}
		& 2 F^{cd} k^a k^b \gamma_{c}^{\ e} \gamma_{d}^{\ f} \nabla_f \nabla_b F_{ae} \\
		= & 2 \left(\gamma^{c}_{\ g} \gamma^{d}_{\ h} F_{cd} \right) \gamma^{ge} \gamma^{hf} k^a k^b \nabla_f \nabla_b F_{ae}\\
		= & 2 \left(\gamma^{c}_{\ g} \gamma^{d}_{\ h} F_{cd} \right) \gamma^{hf} \nabla_f \left(\gamma^{ge} k^a k^b \nabla_b F_{ae} \right) - 2 \left(\gamma^{c}_{\ g} \gamma^{d}_{\ h} F_{cd} \right) \gamma^{hf} \gamma^{ge} k^b \nabla_b F_{ae} \nabla_f k^a \\
		& - 2 \left(\gamma^{c}_{\ g} \gamma^{d}_{\ h} F_{cd} \right) \gamma^{hf} \gamma^{ge} k^a \nabla_b F_{ae} \nabla_f k^b\,.
	\end{split}
\end{equation}
For the first term of Eq. (\ref{hkk2sixteenth}), we have
\begin{equation}
	\begin{split}
		& 2 \left(\gamma^{c}_{\ g} \gamma^{d}_{\ h} F_{cd} \right) \gamma^{hf} \nabla_f \left(\gamma^{ge} k^a k^b \nabla_b F_{ae} \right) \\
		= & 2 \left(F_{\hat{g} \hat{h}} \right) D^{\hat{h}} \left[k^b \nabla_b \left(\gamma^{\hat{g} e} k^a F_{ae} \right) \right] \\
		= & 2 \left(F^{\hat{g}}_{\ \hat{h}} \right)_0 D^{\hat{h}} \left[\mathcal{L}_k \left(k^a F_{a \hat{g}} \right) \right]_1 - 2 \left(F^{\hat{g}}_{\ \hat{h}} \right)_0 D^{\hat{h}} \left[\left(k^a F_{a \hat{f}} \right)_1 \left(\gamma^{f}_{\ b} \nabla_{\hat{g}} k^b \right)_1 \right] \\
		\simeq & 2 \left(F^{\hat{e} \hat{d}} \right)_0 D_{\hat{d}} \left[\mathcal{L}_k \left(k^a F_{a \hat{e}} \right) \right]_1\\
		\sim & \mathcal{O} \left(\epsilon \right)\,.
	\end{split}
\end{equation}
The second term of Eq. (\ref{hkk2sixteenth}) can be simplified as
\begin{equation}
	\begin{split}
		& - 2 \left(\gamma^{c}_{\ g} \gamma^{d}_{\ h} F_{cd} \right) \gamma^{hf} \gamma^{ge} k^b \nabla_b F_{ae} \nabla_f k^a\\
		= & - 2 \left(F^{\hat{g}}_{\ \hat{h}} \right) k^b \nabla_b \left(F_{\hat{m} \hat{g}} \right) \left(\gamma^{\hat{h}f} \nabla_f k^{\hat{m}} \right) + 2 \left(F^{\hat{g}}_{\ \hat{h}} \right) k^b \nabla_b \left(k^a F_{a\hat{g}} \right) \left(l_m \gamma^{\hat{h} f} \nabla_f k^m \right) \\
		& + 2 \left(F^{\hat{g}}_{\ \hat{h}} \right) k^b \nabla_b \left(l^a F_{a \hat{g}} \right) \left(\gamma^{\hat{h}f} k_m \nabla_f k^m \right) \\
		= & - 2 \left(F^{\hat{g}}_{\ \hat{h}} \right)_0 \left[\mathcal{L}_k \left(F_{\hat{m} \hat{g}} \right) \right]_1 \left(\gamma^{\hat{h}f} \nabla_f k^{\hat{m}} \right)_1 + 2 \left(F^{\hat{g}}_{\ \hat{h}} \right)_0 \left(F_{\hat{m} \hat{n}} \right)_0 \left(\gamma^{\hat{n}}_{\ b} \nabla_{\hat{g}} k^b \right)_1 \left(\gamma^{\hat{h}f} \nabla_f k^{\hat{m}} \right)_1 \\
		& + 2 \left(F^{\hat{g}}_{\ \hat{h}} \right)_0 \left(F_{\hat{n} \hat{g}} \right)_0 \left(\gamma^{\hat{n}}_{\ b} \nabla_m k^b \right)_1 \left(\gamma^{\hat{h}f} \nabla_f k^m \right)_1 + 2 \left(F^{\hat{g}}_{\ \hat{h}} \right)_0 \left[\mathcal{L}_k \left(k^a F_{a\hat{g}} \right) \right]_1  \left(l_m \gamma^{\hat{h}f} \nabla_f k^m \right)_1 \\
		& - 2 \left(F^{\hat{g}}_{\ \hat{h}} \right)_0 \left(k^a F_{a\hat{n}} \right)_1 \left(\gamma^{\hat{n}}_{\ b} \nabla_{\hat{g}} k^b \right)_1 \left(l_m \gamma^{\hat{h}f} \nabla_f k^m \right)_1\\
		\simeq & 0\,.
	\end{split}
\end{equation}
The third term of Eq. (\ref{hkk2sixteenth}) can be expressed as
\begin{equation}
	\begin{split}
		& - 2 \left(\gamma^{c}_{\ g} \gamma^{d}_{\ h} F_{cd} \right) \gamma^{hf} \gamma^{ge} k^a \nabla_b F_{ae} \nabla_f k^b\\
		= & - 2 \left(F^{\hat{g}}_{\ \hat{h}} \right) \left(k^a \gamma_{\hat{g}}^{\ e} \gamma^{b}_{\ m} \nabla_b F_{ae} \right) \left(\gamma^{\hat{h}f} \nabla_f k^m \right) + 2 \left(F^{\hat{g}}_{\ \hat{h}} \right) k^b \nabla_b \left(k^a F_{a \hat{g}} \right) \left(l_m \gamma^{hf} \nabla_f k^m \right)\\
		& + 2 \left(F^{\hat{g}}_{\ \hat{h}} \right) \left(\gamma_{\hat{g}}^{\ e} k^a l^b \nabla_b F_{ae} \right) \left(\gamma^{\hat{h}f} k_m \nabla_f k^m \right) \\
		= & - 2 \left(F^{\hat{g}}_{\ \hat{h}} \right)_0 \left[D_m \left(k^a \gamma_{\hat{g}}^{\ e} F_{ae} \right) \right]_1 \left(\gamma^{\hat{h}f} \nabla_f k^m \right)_1 + 2 \left(F^{\hat{g}}_{\ \hat{h}} \right)_0 \left(\gamma_{\hat{g}}^{\ e} F_{ae} \right) \left(\gamma^{b}_{\ m} \nabla_b k^a \right)_1 \left(\gamma^{\hat{h}f} \nabla_f k^m \right)_1 \\
		& + 2 \left(F^{\hat{g}}_{\ \hat{h}} \right)_0 \left[\mathcal{L}_k \left(k^a \gamma_{\hat{g}}^{\ e} F_{ae} \right) \right] _1 \left(l_m \gamma^{\hat{h}f} \nabla_f k^m \right)_1 - 2 \left(F^{\hat{g}}_{\ \hat{h}} \right)_0 \left(\gamma_{n}^{\ e} k^a F_{ae} \right)_1 \left(\gamma^{n}_{\ b} \nabla_{\hat{g}} k^b \right)_1 \left(l_m \gamma^{\hat{h}f} \nabla_f k^m \right)_1 \\
		\simeq & 0\,.
	\end{split}
\end{equation}
So the sixteenth term of Eq. (\ref{hkk2}) under the first-order approximation can be obtained as
\begin{equation}
	2 F^{cd} k^a k^b \gamma_{c}^{\ e} \gamma_{d}^{\ f} \nabla_f \nabla_b F_{ae} \simeq 2 \left(F^{\hat{e} \hat{d}} \right) D_{\hat{d}} \left[\mathcal{L}_k \left(k^a F_{a \hat{e}} \right) \right]
\end{equation}

The seventeenth term of Eq. (\ref{hkk2}) can be further calculated as
\begin{equation}
	\begin{split}
		& 2 F_{a}^{\ c} k^a k^b \gamma_{c}^{\ d} \gamma^{ef} \nabla_f \nabla_b F_{de} \\
		= & 2 \left(k^a F_{ac} \gamma^{cg} \right) \gamma_{h}^{\ f} \nabla_f \left(\gamma_{g}^{\ d} \gamma^{eh} k^b \nabla_b F_{de} \right) - 2 \left(k^a F_{ac} \gamma^{cg} \right)_1 \left(\gamma_{g}^{\ d} \gamma^{eh} F_{de} \right)_0 \left(\gamma_{h}^{\ f} \nabla_f k^b \right)_1 \\
		\simeq & 2 \left(k^a F_{a}^{\ \hat{g}} \right)_1 D^{\hat{h}} \left[\mathcal{L}_k \left(F_{\hat{g} \hat{h}} \right) \right]_1 - 2 \left(k^a F_{a}^{\ \hat{g}} \right)_1 D^{\hat{h}} \left[\left(F_{\hat{f} \hat{h}} \right)_0 \left(\gamma^{\hat{f}}_{\ b} \nabla_{\hat{g}} k^b \right)_1 \right] \\
		& - 2 \left(k^a F_{a}^{\ \hat{g}} \right)_1 D^{\hat{h}} \left[\left(F_{\hat{g} \hat{f}} \right)_0 \left(\gamma^{\hat{f}}_{\ b} \nabla_{\hat{h}} k^b \right)_1 \right] \\
		\simeq & 0\,.
	\end{split}
\end{equation}

The eighteenth term of Eq. (\ref{hkk2}) can be written as
\begin{equation}
	\begin{split}
		& - 2 F_{a}^{\ e} k^a k^b k^c l^d \gamma_{e}^{\ f} \nabla_f \nabla_c F_{bd} \\
		= & - 2 \left(k^a F_{ae} \right) \gamma^{ef} \nabla_f \left(k^b l^d k^c \nabla_c F_{bd} \right) + 2 \left(k^a F_{a\hat{e}} \right)_1 \left(\gamma^{\hat{e}f} \nabla_f k^b \right)_1 \left(l^d k^c \nabla_c F_{bd} \right) \\
		& + 2 \left(k^a F_{ae} \right) \left(\gamma^{ef} \nabla_f l^d \right) \left(k^b k^c \nabla_c F_{bd} \right) + 2 \left(k^a F_{a\hat{e}} \right)_1 \left(\gamma^{\hat{e}f} \nabla_f k^c \right)_1 \left(k^b l^d \nabla_c F_{bd} \right)\\
		\simeq & - 2 \left(k^a F_{a \hat{e}} \right)_1 D^{\hat{e}} \left[\mathcal{L}_k \left(k^b F_{bd} l^d \right) \right]_1 + 2 \left(k^a F_{a\hat{e}} \right) k^c \nabla_c \left(k^b F_{b\hat{h}} \right) \left(\nabla^{\hat{e}} l^{\hat{h}} \right)\\
		& - 2 \left(k^a F_{a \hat{e}} \right) k^c \nabla_c \left(k^{(b} k^{d)} F_{[bd]} \right) \left(l^h \nabla^{\hat{e}} l_h \right) + 2 \left(k^a F_{a \hat{e}} \right)_1 \left[\mathcal{L}_k \left(k^b F_{bd} l^d \right) \right]_1 \left(l_h \gamma^{\hat{e}f} \nabla_f k^h \right)_1\\
		\simeq & 2 \left(k^a F_{a \hat{e}} \right)_1 \left[\mathcal{L}_k \left(k^b F_{b\hat{h}} \right) \right]_1 \left(\nabla^{\hat{e}} l^{\hat{h}} \right) - 2 \left(k^a F_{a \hat{e}} \right)_1 \left(k^b F_{b\hat{c}} \right)_1 \left(\nabla^{\hat{e}} l^h \right) \left(\nabla_h k^{\hat{c}} \right) \\
		\simeq & 0\,.
	\end{split}
\end{equation}

For the nineteenth term of Eq. (\ref{hkk2}), it can be expressed as
\begin{equation}\label{hkk2nineteen}
	\begin{split}
		& - 2 F_{ad} k^a k^b k^c l^d \gamma^{ef} \nabla_f \nabla_c F_{be} \\
		= & - 2 \left(k^a F_{ad} l^d \right) \gamma^{gf} \nabla_f \left(k^b k^c \gamma_{g}^{\ e} \nabla_c F_{be} \right) + 2 \left(k^a F_{ad} l^d \right)  k^c\nabla_c \left(\gamma_{g}^{\ e} F_{be} \right) \left(\gamma^{gf} \nabla_f k^b \right)\\
		& + 2 \left(k^a F_{ad} l^d \right) \left(k^b \gamma_{g}^{\ e} \nabla_c F_{be} \right) \left(\gamma^{gf} \nabla_f k^c \right)\,.
	\end{split}
\end{equation}
For the first term of Eq. (\ref{hkk2nineteen}), it can be calculated as
\begin{equation}
	\begin{split}
		& - 2 \left(k^a F_{ad} l^d \right) \gamma^{gf} \nabla_f \left(k^b k^c \gamma_{g}^{\ e} \nabla_c F_{be} \right)\\
		= & - 2 \left(k^a F_{ad} l^d \right)_0 D^g \left[\mathcal{L}_k \left(k^b F_{be} \gamma^{e}_{\ g} \right) - \left(k^b F_{b\hat{e}} \right)_1 \left(\gamma^{\hat{e}}_{\ c} \nabla_g k^c \right)_1 \right]\\
		\simeq & - 2 \left(k^a F_{ad} l^d \right)_0 D^{\hat{e}} \left[\mathcal{L}_k \left(k^b F_{b\hat{e}} \right) \right]_1\\
		\sim & \mathcal{O} \left(\epsilon \right)\,.
	\end{split}
\end{equation}
The second term of Eq. (\ref{hkk2nineteen}) is
\begin{equation}
	\begin{split}
		& 2 \left(k^a F_{ad} l^d \right)  k^c\nabla_c \left(\gamma_{g}^{\ e} F_{be} \right) \left(\gamma^{gf} \nabla_f k^b \right)\\
		= & 2 \left(k^a F_{ad} l^d \right)_0 \left[\mathcal{L}_k \left(F_{\hat{h} \hat{g}} \right) \right]_1 \left(\gamma^{\hat{g}f} \nabla_f k^{\hat{h}} \right)_1 - 2 \left(k^a F_{ad} l^d \right)_0 \left(F_{\hat{h} \hat{m}} \right)_0 \left(\gamma^{\hat{m}}_{\ c} \nabla_g k^c \right)_1 \left(\gamma^{gf} \nabla_f k^{\hat{h}} \right)_1 \\
		& - 2 \left(k^a F_{ad} l^d \right)_0 \left(F_{\hat{m} \hat{g}} \right)_0 \left(\gamma^{\hat{m}}_{\ c} \nabla_h k^c \right)_1 \left(\gamma^{\hat{g}f} \nabla_f k^h \right)_1 - 2  \left(k^a F_{ad} l^d \right)_0 \left[\mathcal{L}_k \left(k^b F_{b\hat{g}} \right) \right] \left(l_h \gamma^{\hat{g}f} \nabla_f k^h \right)_1 \\
		& + 2 \left(k^a F_{ad} l^d \right)_0 \left(k^b F_{b\hat{m}} \right)_1 \left(\gamma^{\hat{m}}_{\ c} \nabla_g k^c \right)_1 \left(l_h \gamma^{gf} \nabla_f k^h \right)_1\\
		\simeq & 0\,.
	\end{split}
\end{equation}
The third term of Eq. (\ref{hkk2nineteen}) can be expressed as
\begin{equation}
	\begin{split}
		& 2 \left(k^a F_{ad} l^d \right) \left(k^b \gamma_{g}^{\ e} \nabla_c F_{be} \right) \left(\gamma^{gf} \nabla_f k^c \right)\\
		= & 2 \left(k^a F_{ad} l^d \right)_0 D_h \left(k^b F_{b \hat{g}} \right)_1 \left(\gamma^{\hat{g}f} \nabla_f k^h \right)_1 - 2 \left(k^a F_{ad} l^d \right)_0 \left(\gamma^{c}_{\ h} \nabla_c k^b \right)_1 \left(F_{b \hat{g}} \right) \left(\gamma^{\hat{g}f} \nabla_f k^h \right)_1 \\
		& - 2 \left(k^a F_{ad} l^d \right)_0 \left[\mathcal{L}_k \left(k^b F_{b\hat{g}} \right) \right]_1 \left(l_h \gamma^{gf} \nabla_f k^h \right)_1 \\
		& + 2 \left(k^a F_{ad} l^d \right)_0 \left(k^b F_{b \hat{m}} \right)_1 \left(\gamma^{\hat{m}}_{\ c} \nabla_g k^c\right)_1 \left(l_h \gamma^{gf} \nabla_f k^h \right)_1\\
		\simeq & 0\,.
	\end{split}
\end{equation}
Therefore, the result of the nineteenth term of Eq. (\ref{hkk2}) is
\begin{equation}
	\begin{split}
		- 2 F_{ad} k^a k^b k^c l^d \gamma^{ef} \nabla_f \nabla_c F_{be} \simeq - 2 \left(k^a F_{ad} l^d \right) D^{\hat{e}} \left[\mathcal{L}_k \left(k^b F_{b\hat{e}} \right) \right]\,.
	\end{split}
\end{equation}

So the expression of $H_{kk}^{(2)}$ under the linear order approximation can be written as
\begin{equation}
	\begin{split}
		H_{kk}^{(2)} \simeq & - 2 \left[\mathcal{L}_k \left(k^b F_{b}^{\ \hat{e}} \right) \right] \left[D_{\hat{e}} \left(k^a F_{ad} l^d \right) \right]  - 2 \left(k^a F_{ad} l^d \right) D^{\hat{e}} \left[\mathcal{L}_k \left(k^b F_{b\hat{e}} \right) \right] \\
		& + 2 \mathcal{L}_k \left(k^a F_{a\hat{c}} \right) D_{\hat{e}} \left( F^{\hat{c} \hat{e}} \right) + 2 \left(F^{\hat{e} \hat{d}} \right) D_{\hat{d}} \left[\mathcal{L}_k \left(k^a F_{a \hat{e}} \right) \right] \\
		& - 4 \left(k^a F_{a}^{\ \hat{e}} \right) \left[ \mathcal{L}_k^2 \left(F_{\hat{e}d} l^d \right) \right] + 4 \left(k^a F_{a}^{\ \hat{e}} \right) \left[\mathcal{L}_k^2 \left(F_{\hat{e}d} l^d \right) \right] \\
		& - 2 \left(k^a F_{ae} l^e \right) \left(k^b F_{bf} l^f \right) \left(R_{kk} \right) + 4 \left(k^a F_{ae} l^e \right) \left[\mathcal{L}_k^2 \left(k^b F_{bf} l^f \right) \right] \\
		& + 2 \left(l^d F_{d}^{\ \hat{e}} \right) \left[\mathcal{L}_k^2 \left(k^a F_{a \hat{e}} \right) \right] + 2 \left(k^a F_{a \hat{e}} \right) \left[\mathcal{L}_k^2 \left(l^d F_{d}^{\ \hat{e}} \right) \right] \\
		& + 4 \left[\mathcal{L}_k \left(k^a F_{a \hat{e}} \right) \right] \left[\mathcal{L}_k \left(l^d F_{d}^{\ \hat{e}} \right) \right]\,.
	\end{split}
\end{equation}

\section{Simplification of $H_{kk}^{(3)}$ under the linear order approximation}\label{appd}
In the final appendix, we will calculate Eq. (\ref{ohkk3}) under the first-order approximation. After duplicated indexes in every term of the equation as well, Eq. (\ref{ohkk3}) can be further written as
\begin{equation}\label{hkk3}
	\begin{split}
		H_{kk}^{(3)} = & - 2 F_{a}^{\ c} F^{de} k^a k^b R_{bfgh} \gamma_{c}^{\ f} \gamma_{d}^{\ g} \gamma_{e}^{\ h} + 4 F_{a}^{\ e} F_{d}^{\ f} k^a k^b k^c l^d R_{bgch} \gamma_{e}^{\ g} \gamma_{f}^{\ h} \\
		& + 4 F_{a}^{\ e} F_{b}^{\ f} k^a k^b k^c l^d R_{cgdh} \gamma_{e}^{\ g} \gamma_{f}^{\ h} + 4 F_{ae} F_{b}^{\ g} k^a k^b k^c k^d l^e l^f R_{cfdh} \gamma_{g}^{\ h} \\
		& - 4 F_{a}^{\ e} k^a k^b k^c l^d \gamma_{e}^{\ f} \nabla_c \nabla_f F_{bd} - 4 F_{a}^{\ e} k^a k^b k^c l^d \gamma_{e}^{\ f} \nabla_f \nabla_c F_{bd} \\
		& - 4 F_{ad} k^a k^b k^c l^d \gamma^{ef} \nabla_c \nabla_f F_{be} - 4 F_{ad} k^a k^b k^c l^d \gamma^{ef} \nabla_f \nabla_c F_{be}\\
		& + 8 k^a k^b k^c k^d l^e l^f \nabla_b F_{ae} \nabla_d F_{cf} + 8 F_{ae} k^a k^b k^c k^d l^e l^f \nabla_d \nabla_c F_{bf}\\
		& + 4 F_{a}^{\ c} k^a k^b \gamma_{c}^{\ d} \gamma^{ef} \nabla_d \nabla_f F_{be} + 4 F_{a}^{\ c} k^a k^b \gamma_{c}^{\ d} \gamma^{ef} \nabla_f \nabla_d F_{be}\\
		& + 4 k^a k^b \gamma^{cd} \gamma^{ef} \nabla_d F_{bf} \nabla_e F_{ac} - 8 k^a k^b k^c l^d \gamma^{ef} \nabla_c F_{bf} \nabla_e F_{ad} \\
		& + 4 k^a k^b \gamma^{cd} \gamma^{ef} \nabla_d F_{ac} \nabla_f F_{be} - 8 k^a k^b k^c l^d \gamma^{ef} \nabla_b F_{ad} \nabla_f F_{ce}\,.
	\end{split}
\end{equation}
The first term of Eq. (\ref{hkk3}) is
\begin{equation}
	\begin{split}
		- 2 F_{a}^{\ c} F^{de} k^a k^b R_{bfgh} \gamma_{c}^{\ f} \gamma_{d}^{\ g} \gamma_{e}^{\ h} = - 2 \left(k^a F_{a}^{\ \hat{c}} \right)_1 \left(F^{\hat{d} \hat{e}} \right)_0 \left(k^b R_{b \hat{c} \hat{d} \hat{e}} \right)_1 \simeq  0\,.
	\end{split}
\end{equation}
The second term of Eq. (\ref{hkk3}) can be calculated as
\begin{equation}
	\begin{split}
		4 F_{a}^{\ e} F_{d}^{\ f} k^a k^b k^c l^d R_{bgch} \gamma_{e}^{\ g} \gamma_{f}^{\ h} = 4 \left(k^a F_{a}^{\ \hat{e}} \right)_1 \left(l^d F_{d}^{\ \hat{f}} \right)_0 \left(k^b k^c R_{b\hat{e}c\hat{f}} \right)_1 \simeq 0\,.
	\end{split}
\end{equation}
The third term of Eq. (\ref{hkk3}) can be written as
\begin{equation}
	\begin{split}
		4 F_{a}^{\ e} F_{b}^{\ f} k^a k^b k^c l^d R_{cgdh} \gamma_{e}^{\ g} \gamma_{f}^{\ h} = \left(k^a F_{a}^{\ \hat{e}} \right)_1 \left(k^b F_{b}^{\ \hat{f}} \right)_1 \left(k^c l^d R_{c \hat{e} d \hat{f}} \right)_0 \simeq 0\,.
	\end{split}
\end{equation}
The fourth term of Eq. (\ref{hkk3}) is expressed as
\begin{equation}
	\begin{split}
		4 F_{ae} F_{b}^{\ g} k^a k^b k^c k^d l^e l^f R_{cfdh} \gamma_{g}^{\ h} = \left(k^a F_{ae} l^e \right)_0 \left(k^b F_{b}^{\ \hat{g}} \right)_1 \left(k^c l^f k^d R_{cfd \hat{g}} \right)_1 \simeq 0\,.
	\end{split}
\end{equation}
According to the relation (\ref{riemanntensor}), the fifth and sixth terms of Eq. (\ref{hkk3}) can be written as
\begin{equation}\label{fifsixterm}
	\begin{split}
		& - 4 F_{a}^{\ e} k^a k^b k^c l^d \gamma_{e}^{\ f} \nabla_c \nabla_f F_{bd} - 4 F_{a}^{\ e} k^a k^b k^c l^d \gamma_{e}^{\ f} \nabla_f \nabla_c F_{bd}\\
		= & - 8 F_{ae} k^a k^b k^c l^d \gamma^{ef} \nabla_f \nabla_c F_{bd} - 4 F_{ae} k^a k^b k^c l^d \gamma^{ef} R_{cfb}^{\ \ \ g} F_{gd}\\
		& - 4 F_{ae} k^a k^b k^c l^d \gamma^{ef} R_{cfd}^{\ \ \ g} F_{bg}\,.
	\end{split}
\end{equation}
For the first term of Eq. (\ref{fifsixterm}), it is further calculated as
\begin{equation}
	\begin{split}
		& - 8 F_{ae} k^a k^b k^c l^d \gamma^{ef} \nabla_f \nabla_c F_{bd}\\
		= & - 8 \left(k^a F_{a \hat{e}} \right)_1 D^e \left[\mathcal{L}_k \left(k^b l^d F_{bd} \right) \right]_1 +  8 \left(k^a F_{a\hat{e}} \right)_1 \left(\gamma^{\hat{e} f} \nabla_f k^c \right)_1 \left(k^b l^d \nabla_c F_{bd} \right) \\
		& +  8 \left(k^a F_{a\hat{e}} \right)_1 \left(\gamma^{\hat{e} f} \nabla_f k^b \right)_1 \left(l^d k^c \nabla_c F_{bd} \right) +  8 \left(k^a F_{a\hat{e}} \right) \left(k^c k^b \nabla_c F_{bd} \right) \left(\nabla^{\hat{e}} l^d \right) \\
		\simeq & 8 \left(k^a F_{a\hat{e}} \right)_1 \left[\mathcal{L}_k \left(k^b F_{b \hat{g}} \right)  \right]_1 \left(\nabla^{\hat{e}} l^{\hat{g}} \right) - 8 \left(k^a F_{a\hat{e}} \right)_1 \left(k^b F_{b \hat{d}} \right)_1 \left(\gamma^{\hat{d}}_{\ c} \nabla_g k^c \right)_1 \left(\nabla^{\hat{e}} l^g \right)\\
		& -  8 \left(k^a F_{a \hat{e}} \right) k^c \nabla_c \left( k^{(b}  k^{d)} F_{[bd]} \right) \left(l^g \nabla^{\hat{e}} l_g \right) -  8 \left(k^a F_{a\hat{e}} \right)_1 \left[\mathcal{L}_k \left( k^b F_{bd} l^d \right) \right]_1 \left(k^g \nabla^{\hat{e}} l_g \right)\\
		\simeq & 0\,.
	\end{split}
\end{equation}
The second term of Eq. (\ref{fifsixterm}) is
\begin{equation}
	\begin{split}
		& - 4 F_{ae} k^a k^b k^c l^d \gamma^{ef} R_{cfb}^{\ \ \ g} F_{gd} \\
		= &  - 4 \left(k^a F_{a}^{\ \hat{e}} \right)_1 \left(k^b k^c R_{c \hat{e} b \hat{f}} \right)_1 \left(F^{\hat{f}}_{\ d} l^d \right)_0 + 4 \left(k^a F_{a}^{\ \hat{e}} \right)_1 \left(k^c k^b l^h R_{c \hat{e} b h} \right)_1 \left(k^g F_{gd} l^d \right)_0 \\
		& + 4 \left(k^a F_{a}^{\ \hat{e}} \right) \left(k^c k^{(b} k^{h)} R_{c \hat{e} [b h]} \right) \left(l^{(g} l^{d)} F_{[gd]} \right) \\
		\simeq & 0\,.
	\end{split}
\end{equation}
The third term of Eq. (\ref{fifsixterm}) can be calculated as
\begin{equation}
	\begin{split}
		& - 4 F_{ae} k^a k^b k^c l^d \gamma^{ef} R_{cfd}^{\ \ \ g} F_{bg}\\
		= & - 4 \left(k^a F_{a}^{\ \hat{e}}  \right)_1 \left(k^c l^d R_{c \hat{e} d \hat{f}} \right)_0 \left(k^b F_{b}^{\ \hat{f}} \right)_1 + 4 \left(k^a F_{a}^{\ \hat{e}}  \right)_1 \left(k^c l^d l^h R_{c \hat{e} d h} \right)_0 \left(k^{(b} k^{g)} F_{[bg]} \right) \\
		& + 4 \left(k^a F_{a}^{\ \hat{e}} \right)_1 \left(k^c l^d k^h R_{c \hat{e} d h} \right)_1 \left(k^b l^g F_{bg} \right)_0\\
		\simeq & 0\,.
	\end{split}
\end{equation}
So the result of the fifth and sixth terms of Eq. (\ref{hkk3}) under the first-order approximation is
\begin{equation}
	\begin{split}
		& - 4 F_{a}^{\ e} k^a k^b k^c l^d \gamma_{e}^{\ f} \nabla_c \nabla_f F_{bd} - 4 F_{a}^{\ e} k^a k^b k^c l^d \gamma_{e}^{\ f} \nabla_f \nabla_c F_{bd} \simeq 0\,.
	\end{split}
\end{equation}
Utilizing the relation (\ref{riemanntensor}), the seventh and eighth terms of Eq. (\ref{hkk3}) can be combined as
\begin{equation}\label{seveigterms}
	\begin{split}
		& - 4 F_{ad} k^a k^b k^c l^d \gamma^{ef} \nabla_c \nabla_f F_{be} - 4 F_{ad} k^a k^b k^c l^d \gamma^{ef} \nabla_f \nabla_c F_{be}\\
		= & - 8 F_{ad} k^a k^b k^c l^d \gamma^{ef} \nabla_f \nabla_c F_{be} - 4 F_{ad} k^a k^b k^c l^d \gamma^{ef} R_{cfb}^{\ \ \ g} F_{ge} \\
		& - 4 F_{ad} k^a k^b k^c l^d \gamma^{ef} R_{cfe}^{\ \ \ g} F_{bg}\,.
	\end{split}
\end{equation}
The first term of Eq. (\ref{seveigterms}) is given as
\begin{equation}\label{hkk3seeifirstterm}
	\begin{split}
		& - 8 F_{ad} k^a k^b k^c l^d \gamma^{ef} \nabla_f \nabla_c F_{be} \\
		= & - 8 \left(k^a F_{ad} l^d \right) \gamma^{gf} \nabla_f \left(\gamma_{g}^{\ e} k^b k^c \nabla_c F_{be} \right) + 8 \left(k^a F_{ad} l^d \right) \left(\gamma^{ef} \nabla_f k^b \right) k^c \nabla_c F_{be} \\
		& + 8 \left(k^a F_{ad} l^d \right) \left(\gamma^{ef} \nabla_f k^c \right) k^b \nabla_c F_{be}\,.
	\end{split}
\end{equation}
The first term of Eq. (\ref{hkk3seeifirstterm}) is
\begin{equation}
	\begin{split}
		& - 8 \left(k^a F_{ad} l^d \right) \gamma^{gf} \nabla_f \left(\gamma_{g}^{\ e} k^b k^c \nabla_c F_{be} \right)\\
		= & - 8 \left(k^a F_{ad} l^d \right)_0 D^{\hat{e}} \left[\mathcal{L}_k \left(k^b F_{b \hat{e}} \right) \right]_1 + 8 \left(k^a F_{ad} l^d \right)_0 \gamma^{gf} \nabla_f \left[\left(k^b F_{be} \gamma_{h}^{\ e} \right)_1 \left(\gamma^{h}_{\ c} \nabla_g k^c \right)_1 \right]\\
		\simeq & - 8 \left(k^a F_{ad} l^d \right)_0 D^{\hat{e}} \left[\mathcal{L}_k \left(k^b F_{b \hat{e}} \right) \right]_1\\
		\sim & \mathcal{O} \left(\epsilon \right)\,.
	\end{split}
\end{equation}
The second term of Eq. (\ref{hkk3seeifirstterm}) is given as
\begin{equation}
	\begin{split}
		& 8 \left(k^a F_{ad} l^d \right) \left(\gamma^{ef} \nabla_f k^b \right) k^c \nabla_c F_{be}\\
		= & 8 \left(k^a F_{ad} l^d \right)_0 \left(\gamma^{\hat{m}g} \nabla^{\hat{e}} k_g \right)_1 \left[\mathcal{L}_k \left(F_{\hat{m} \hat{e}} \right) \right]_1 - 8 \left(k^a F_{ad} l^d \right)_0 \left(\gamma^{\hat{n}g} \nabla_f k_g \right)_1 \left(F_{\hat{n} \hat{c}} \right)_0 \left(\gamma^{mf} \nabla_m k^{\hat{c}} \right)_1 \\
		& - 8 \left(k^a F_{ad} l^d \right)_0 \left(\gamma^{\hat{m}f} \nabla_f k_g \right)_1 \left(\gamma^{ng} \nabla_n k^{\hat{c}} \right)_1 \left( F_{\hat{c} \hat{m}} \right)_0 - 8 \left(k^a F_{ad} l^d \right)_0 \left(l^g \gamma^{\hat{h}f} \nabla_f k_g \right)_1 \left[\mathcal{L}_k \left(k^b F_{b\hat{h}} \right) \right]_1\\
		& + 8 \left(k^a F_{ad} l^d \right)_0 \left(\gamma^{hf} l^g \nabla_f k_g \right)_1 \left(\gamma^{\hat{m}}_{\ c} \nabla_h k^c \right)_1 \left(k^b F_{b\hat{m}} \right)_1\\
		\simeq & 0\,.
	\end{split}
\end{equation}
The third term of Eq. (\ref{hkk3seeifirstterm}) is calculated as
\begin{equation}
	\begin{split}
		& 8 \left(k^a F_{ad} l^d \right) \left(\gamma^{ef} \nabla_f k^c \right) k^b \nabla_c F_{be}\\
		= & 8 \left(k^a F_{ad} l^d \right)_0 \left(\gamma^{\hat{h}f} \nabla_f k^g \right)_1 D_g \left(k^b F_{b\hat{h}} \right)_1 - 8 \left(k^a F_{ad} l^d \right)_0 \left(\gamma^{hf} \nabla_f k^g \right)_1 \left(F_{b \hat{h}} \right) \left(\gamma_{g}^{\ c} \nabla_c k^b \right)_1\\
		& - 8 \left(k^a F_{ad} l^d \right)_0 \left(l_g \gamma^{\hat{h}f} \nabla_f k^g \right)_1 \left[\mathcal{L}_k \left(k^b F_{b\hat{h}} \right) \right]_1 + 8 \left(k^a F_{ad} l^d \right)_0 \left(l_g \gamma^{hf} \nabla_f k^g \right)_1 \left(k^b F_{b \hat{e}} \right)_1 \left(\gamma_{c}^{\ \hat{e}} \nabla_h k^c \right)_1\\
		\simeq & 0\,.
	\end{split}
\end{equation}
So the first term of Eq. (\ref{seveigterms}) under the linear order approximation can be finally expressed as
\begin{equation}
	\begin{split}
		- 8 F_{ad} k^a k^b k^c l^d \gamma^{ef} \nabla_f \nabla_c F_{be}\simeq - 8 \left(k^a F_{ad} l^d \right) D^{\hat{e}} \left[\mathcal{L}_k \left(k^b F_{b \hat{e}} \right) \right]\,.
	\end{split}
\end{equation}
The second term of Eq. (\ref{seveigterms}) is
\begin{equation}
	\begin{split}
		& - 4 F_{ad} k^a k^b k^c l^d \gamma^{ef} R_{cfb}^{\ \ \ g} F_{ge} \\
		= & - 4 \left(k^c F_{cd} l^d \right)_0 \left(F^{[\hat{e} \hat{f}]} \right)_0 \left(k^a k^b R_{a (\hat{e} \hat{f}) b} \right)_1 + 4 \left(k^d F_{de} l^e \right) \left(l^f F_{f}^{\ \hat{g}} \right) \left(k^a k^{(b} k^{c)} R_{a \hat{g} [b c]} \right)\\
		& + 4 \left(k^c F_{cf} l^f \right)_0 \left(k^d F_{d}^{\ \hat{g}} \right)_1 \left(k^a k^b l^e R_{a \hat{g}be} \right)_1\\
		\simeq & 0\,.
	\end{split}
\end{equation}
The third term of Eq. (\ref{seveigterms}) can be written as
\begin{equation}
	\begin{split}
		& - 4 F_{ad} k^a k^b k^c l^d \gamma^{ef} R_{cfe}^{\ \ \ g} F_{bg} \\
		= &- 4 \left(k^c F_{ce} l^e \right)_0 \left(k^d F_{df} l^f \right)_0 \left(k^a k^b R_{ab} \right)_1 + 4 \left(k^b F_{bd} l^d \right)_0 \left(k^c F_{c}^{\ \hat{g}}  \right)_1 \left(k^a R_{a \hat{g}} \right)_1\\
		\simeq & - 4 \left(k^c F_{ce} l^e \right)_0 \left(k^d F_{df} l^f \right)_0 \left(R_{kk} \right)_1\\
		\sim & \mathcal{O} \left(\epsilon\right)\,.
	\end{split}
\end{equation}
The result of the seventh and eighth terms of Eq. (\ref{hkk3}) is
\begin{equation}
	\begin{split}
		& - 4 F_{ad} k^a k^b k^c l^d \gamma^{ef} \nabla_c \nabla_f F_{be} - 4 F_{ad} k^a k^b k^c l^d \gamma^{ef} \nabla_f \nabla_c F_{be}\\
		\simeq & - 8 \left(k^a F_{ad} l^d \right) D^{\hat{e}} \left[\mathcal{L}_k \left(k^b F_{b \hat{e}} \right) \right] - 4 \left(k^c F_{ce} l^e \right) \left(k^d F_{df} l^f \right) \left(R_{kk} \right)\,.
	\end{split}
\end{equation}
The ninth term of Eq. (\ref{hkk3}) can be written as
\begin{equation}
	\begin{split}
		8 k^a k^b k^c k^d l^e l^f \nabla_b F_{ae} \nabla_d F_{cf} = 8 \left[\mathcal{L}_k  \left(k^a F_{ae} l^e \right) \right]_1 \left[\mathcal{L}_k \left(k^c F_{cf} l^f \right) \right]_1 \simeq 0\,.
	\end{split}
\end{equation}
For the tenth term of Eq. (\ref{hkk3}), we have
\begin{equation}
	\begin{split}
		8 F_{ae} k^a k^b k^c k^d l^e l^f \nabla_d \nabla_c F_{bf} = 8 \left(k^a F_{ae} l^e \right)_0 \left[\mathcal{L}_k^2 \left(k^b F_{bf} l^f \right)  \right]_1 \sim \mathcal{O} \left(\epsilon \right)\,.
	\end{split}
\end{equation}
Using the relation (\ref{riemanntensor}) again, the eleventh and twelfth terms of Eq. (\ref{hkk3}) can be expressed as
\begin{equation}\label{eletwelterms}
	\begin{split}
		& 4 F_{a}^{\ c} k^a k^b \gamma_{c}^{\ d} \gamma^{ef} \nabla_d \nabla_f F_{be} +  4 F_{a}^{\ c} k^a k^b \gamma_{c}^{\ d} \gamma^{ef} \nabla_f \nabla_d F_{be}\\
		= & 8 k^a F_{ac} \gamma^{cd} \gamma^{ef} k^b \nabla_f \nabla_d F_{be} + 4 F_{ac} k^a k^b \gamma^{cd} \gamma^{ef} R_{dfb}^{\ \ \ g} F_{ge} \\
		& + 4 F_{ac} k^a k^b \gamma^{cd} \gamma^{ef} R_{dfe}^{\ \ \ g} F_{bg}\,.
	\end{split}
\end{equation}
For the first term of Eq. (\ref{eletwelterms}), we have
\begin{equation}
	\begin{split}
		& 8 k^a F_{ac} \gamma^{cd} \gamma^{ef} k^b \nabla_f \nabla_d F_{be} \\
		= & 8 \left(k^a F_{a}^{\ \hat{g}} \right) \gamma_{h}^{\ f}  \nabla_f \left(\gamma_{\hat{g}}^{\ d} \gamma^{eh}  k^b \nabla_d F_{be} \right) - 8 \left(k^a F_{a}^{\ \hat{g}} \right)_1 \left(\gamma_{h}^{\ f}  \nabla_f  k^b \right)_1 \left(\gamma_{\hat{g}}^{\ d} \gamma^{eh}  \nabla_d F_{be} \right) \\
		\simeq & 8 \left(k^a F_{a}^{\ \hat{g}} \right)_1 D_{\hat{h}} \left[D_{\hat{g}} \left(k^b F_{b}^{\ \hat{h}} \right)_1 \right] - 8 \left(k^a F_{a}^{\ \hat{g}} \right)_1 D_{\hat{h}} \left[\left(F_{b}^{\ \hat{h}} \right) \left(\gamma_{\hat{g}}^{\ d} \nabla_d k^b \right)_1 \right]\\
		\simeq & 0\,.
	\end{split}
\end{equation}
The second term of Eq. (\ref{eletwelterms}) is
\begin{equation}
	\begin{split}
		& 4 F_{ac} k^a k^b \gamma^{cd} \gamma^{ef} R_{dfb}^{\ \ \ g} F_{ge} \\
		= & 4 \left(k^a F_{a}^{\ \hat{c}} \right)_1 \left(k^b R_{\hat{c} \hat{e} b \hat{g}} \right)_1 \left(F^{\hat{g} \hat{e}} \right)_0 - 4 \left(k^a F_{a}^{\ \hat{c}} \right)_1 \left(k^b l^h R_{\hat{c} \hat{e} b h} \right)_0 \left(k^g F_{g}^{\ \hat{e}} \right)_1\\
		& - 4 \left(k^a F_{a}^{\ \hat{c}} \right)_1 \left(k^{(b} k^{h)} R_{\hat{c} \hat{e} [b h]} \right) \left(l^g F_{g}^{\ \hat{e}} \right)\\
		\simeq & 0\,.
	\end{split}
\end{equation}
The third term of Eq. (\ref{eletwelterms}) can be calculated as
\begin{equation}
	\begin{split}
		& 4 F_{ac} k^a k^b \gamma^{cd} \gamma^{ef} R_{dfe}^{\ \ \ g} F_{bg} \\
		= & - 4 \left(k^a F_{a}^{\ \hat{c}} \right)_1 \left(R_{\hat{c} \hat{g}} \right)_0 \left(k^b F_{b}^{\ \hat{g}} \right)_1 + 4 \left(k^a F_{a}^{\ \hat{c}} \right) \left(l^h R_{\hat{c} h} \right) \left(k^{(b} k^{g)} F_{[bg]} \right)\\
		& + 4 \left(k^a F_{a}^{\ \hat{c}} \right)_1 \left(k^h R_{\hat{c} h} \right)_1 \left(k^b F_{bg} l^g \right)_0\\
		\simeq & 0\,.
	\end{split}
\end{equation}
After combining the eleventh and twelfth terms of Eq. (\ref{hkk3}), the result under the first-order approximation can be obtained as
\begin{equation}
	\begin{split}
		4 F_{a}^{\ c} k^a k^b \gamma_{c}^{\ d} \gamma^{ef} \nabla_d \nabla_f F_{be} +  4 F_{a}^{\ c} k^a k^b \gamma_{c}^{\ d} \gamma^{ef} \nabla_f \nabla_d F_{be} \simeq 0\,.
	\end{split}
\end{equation}
The thirteenth term of Eq. (\ref{hkk3}) can be expressed as
\begin{equation}
	\begin{split}
		& 4 k^a k^b \gamma^{cd} \gamma^{ef} \nabla_d F_{bf} \nabla_e F_{ac}\\
		= & 4 \left[D^{\hat{g}} \left(k^b F_{b\hat{h}} \right)_1 - \left(\gamma^{\hat{g}d} \nabla_d k^b \right)_1 \left(F_{b\hat{h}} \right) \right] \left[D^{\hat{h}} \left( k^a F_{a\hat{g}} \right)_1 - \left(\gamma^{e\hat{h}} \nabla_e k^a  \right)_1 \left(F_{a\hat{g}} \right) \right]\\
		\simeq & 0\,.
	\end{split}
\end{equation}
The fourteenth term of Eq. (\ref{hkk3}) is
\begin{equation}\label{fourteenthtermofhkk3}
	\begin{split}
		& - 8 k^a k^b k^c l^d \gamma^{ef} \nabla_c F_{bf} \nabla_e F_{ad}\\
		= & - 8 k^c \nabla_c \left(\gamma_{g}^{\ f} k^b F_{bf} \right) \gamma^{ge} \nabla_e \left(k^a F_{ad} l^d \right) + 8 k^c \nabla_c \left(\gamma_{g}^{\ f} k^b F_{bf} \right) \gamma^{ge} l^d F_{ad} \nabla_e k^a \\
		& + 8 k^c \nabla_c \left(\gamma_{g}^{\ f} k^b F_{bf} \right) \gamma^{ge} k^a F_{ad} \nabla_e l^d\,.
	\end{split}
\end{equation}
The first term of Eq. (\ref{fourteenthtermofhkk3}) can be expressed as
\begin{equation}
	\begin{split}
		& - 8 k^c \nabla_c \left(\gamma_{g}^{\ f} k^b F_{bf} \right) \gamma^{ge} \nabla_e \left(k^a F_{ad} l^d \right) \\
		= & - 8 \left[\mathcal{L}_k \left(k^b F_{b\hat{g}} \right) \right]_1 D^{\hat{g}} \left(k^a F_{ad} l^d \right)_0 + 8 \left(k^b F_{b\hat{h}} \right)_1 \left(\gamma^{\hat{h}}_{\ c} \nabla_g k^c \right)_1 D^g \left(k^a F_{ad} l^d \right)_0 \\
		= & - 8 \left[\mathcal{L}_k \left(k^b F_{b \hat{c}} \right) \right]_1 D^{\hat{c}} \left(k^a F_{ad} l^d \right)_0 \\
		\sim & \mathcal{O} \left(\epsilon \right)\,.
	\end{split}
\end{equation}
The second term of Eq. (\ref{fourteenthtermofhkk3}) is
\begin{equation}
	\begin{split}
		& 8 k^c \nabla_c \left(\gamma_{g}^{\ f} k^b F_{bf} \right) \gamma^{ge} l^d F_{ad} \nabla_e k^a\\
		= & 8 \left[\mathcal{L}_k \left(k^b F_{b\hat{g}} \right) \right]_1 \left(\gamma^{\hat{g}e} \nabla_e k^a \right)_1 \left(l^d F_{ad} \right) - 8 \left(k^b F_{b\hat{h}} \right)_1 \left(\gamma^{\hat{h}}_{\ c} \nabla_g k^c \right)_1 \left(\gamma^{ge} \nabla_e k^a \right)_1 \left(l^d F_{ad} \right)\\
		\simeq & 0\,.
	\end{split}
\end{equation}
The third term of Eq. (\ref{fourteenthtermofhkk3}) is
\begin{equation}
	\begin{split}
		& 8 k^c \nabla_c \left(\gamma_{g}^{\ f} k^b F_{bf} \right) \gamma^{ge} k^a F_{ad} \nabla_e l^d\\
		= & 8 k^c \nabla_c \left(\gamma_{g}^{\ f} k^b F_{bf} \right) \left(k^a F_{ad} \gamma^{hd} \right) \left(\gamma^{ge} \nabla_e l_h \right) + 8 k^c \nabla_c \left(\gamma_{g}^{\ f} k^b F_{bf} \right) \left(k^a F_{ad} l^d \right)\left(\gamma^{ge} l_h \nabla_e k^h \right) \\
		& - 8 k^c \nabla_c \left(\gamma_{g}^{\ f} k^b F_{bf} \right) \left(k^{(a} k^{d)} F_{[ad]}  \right) \left(\gamma^{ge} l^h \nabla_e l_h \right)\\
		= &  8 \left[\mathcal{L}_k \left(k^b F_{b\hat{g}} \right) \right]_1 \left(k^a F_{a}^{\ \hat{h}} \right)_1 \left(\gamma^{\hat{g}e} \nabla_e l_{\hat{h}} \right) - 8 \left(k^b F_{b\hat{f}} \right) \left(\gamma_{c}^{\ \hat{f}} \nabla_g k^c\right)_1 \left(k^a F_{a}^{\ \hat{h}} \right)_1 \left(\gamma^{ge} \nabla_e l_{\hat{h}} \right) \\
		& + 8 \left[\mathcal{L}_k \left(k^b F_{b\hat{g}} \right) \right]_1 \left(k^a F_{ad} l^d \right)_0 \left(l_h \gamma^{\hat{g}e} \nabla_e k^h \right)_1 - 8 \left(k^b F_{b\hat{f}} \right)_1 \left(\gamma_{c}^{\ \hat{f}} \nabla_g k^c\right)_1 \left(k^a F_{ad} l^d \right)_0 \left(l_h \gamma^{ge} \nabla_e k^h \right)_1\\
		\simeq & 0\,.
	\end{split}
\end{equation}
So the result of the fourteenth term of Eq. (\ref{hkk3}) can be obtained as
\begin{equation}
	- 8 k^a k^b k^c l^d \gamma^{ef} \nabla_c F_{bf} \nabla_e F_{ad} \simeq  - 8 \left[\mathcal{L}_k \left(k^b F_{b \hat{c}} \right) \right] D^{\hat{c}} \left(k^a F_{ad} l^d \right)\,.
\end{equation}

The fifteenth term of Eq. (\ref{hkk3}) is further calculated as
\begin{equation}
	\begin{split}
		& 4 k^a k^b \gamma^{cd} \gamma^{ef} \nabla_d F_{ac} \nabla_f F_{be}\\
		= & 4 \left[D^{\hat{g}} \left(k^a F_{a\hat{g}} \right)_1 - \left(\gamma^{\hat{g}d} \nabla_d k^a \right)_1 \left(F_{a\hat{g}} \right) \right] \left[D^{\hat{h}} \left(k^b F_{b\hat{h}} \right)_1 - \left(\gamma^{\hat{h}f} \nabla_f k^b \right)_1 \left(F_{b\hat{h}} \right) \right]\\
		\simeq &  0\,.
	\end{split}
\end{equation}

The sixteenth term of Eq. (\ref{hkk3}) is
\begin{equation}
	\begin{split}
		& - 8 k^a k^b k^c l^d \gamma^{ef} \nabla_b F_{ad} \nabla_f F_{ce} \\
		= & - 8 \left[\mathcal{L}_k \left(k^a F_{ad} l^d \right) \right]_1 \left[D^{\hat{g}} \left(k^c F_{c \hat{g}} \right)_1 - \left(\gamma^{\hat{g}f} \nabla_f k^c \right)_1 \left(F_{c \hat{g}} \right) \right]\\
		\simeq & 0\,.
	\end{split}
\end{equation}

After retaining the first-order terms and ignoring the high-order term as well, the result of $H_{kk}^{(3)}$ under the first-order approximation can be expressed as
\begin{equation}
	\begin{split}
		H_{kk}^{(3)} \simeq & - 8 \left(k^a F_{ad} l^d \right) D^{\hat{e}} \left[\mathcal{L}_k \left(k^b F_{b \hat{e}} \right) \right] - 8 \mathcal{L}_k \left(k^b F_{b \hat{c}} \right) D^{\hat{c}} \left(k^a F_{ad} l^d \right) \\
		& - 4 \left(k^c F_{ce} l^e \right) \left(k^d F_{df} l^f \right) \left(R_{kk} \right) + 8 \left(k^a F_{ae} l^e \right) \left[\mathcal{L}_k^2 \left(k^b F_{bf} l^f \right) \right]\,.
	\end{split}
\end{equation}

\end{document}